\documentclass[journal]{IEEEtran}
\usepackage{bbding}
\usepackage{threeparttable}
\usepackage{booktabs}
\usepackage{multirow}
\usepackage{makecell}
\usepackage[T1]{fontenc}
\usepackage{graphicx}
\usepackage{bm}
\usepackage[noend]{algpseudocode}
\usepackage{algorithmicx,algorithm}
\usepackage{threeparttable}
\usepackage[sort]{cite}
\usepackage{verbatim}
\usepackage{hyperref}
\hypersetup{hidelinks}

\def\equationautorefname#1#2\null{
	Eq. (#2\null)
}

\ifCLASSOPTIONcompsoc
  \usepackage[caption=false,font=normalsize,labelfont=sf,textfont=sf]{subfig}
\else
  \usepackage[caption=false,font=footnotesize]{subfig}
\fi

\usepackage{amsmath}
\usepackage[cmintegrals]{newtxmath}
\usepackage{stfloats}
\usepackage{xspace}
\usepackage{xcolor}

\newcommand{\method}{CASTLE\xspace}

\begin{document}
%
\title{Boosting Cross-Domain Speech Recognition with Self-Supervision}
%
%
%

\author{Han~Zhu,~\IEEEmembership{Student~Member,~IEEE,}
        Gaofeng~Cheng,~\IEEEmembership{Member,~IEEE,}
        Jindong~Wang,
        Wenxin~Hou,
        Pengyuan~Zhang,~\IEEEmembership{Member,~IEEE,}
        Yonghong~Yan,~\IEEEmembership{Member,~IEEE,}
\thanks{H. Zhu, G. Cheng, P. Zhang and Y. Yan are with the Key Laboratory of Speech Acoustics and Content Understanding, Institute of Acoustics, Chinese Academy of Sciences, Beijing, China, and also with the University of Chinese Academy of Sciences, Beijing, China (e-mail:zhuhan@hccl.ioa.ac.cn;chenggaofeng@hccl.ioa.ac.cn
;zhangpengyuan@hccl.ioa.ac.cn;yanyonghong@hccl.ioa.ac.cn).}
\thanks{J. Wang (jindong.wang@microsoft.com) is with Microsoft Research Asia, China.}
\thanks{W. Hou (wenxinhou@microsoft.com) is with Microsoft STCA, China.}
}%


%
%

\markboth{Journal of \LaTeX\ Class Files,~Vol.~14, No.~8, August~2015}%
{Shell \MakeLowercase{\textit{et al.}}: Bare Demo of IEEEtran.cls for IEEE Journals}
%



\maketitle
\begin{abstract}

The cross-domain performance of automatic speech recognition (ASR) could be severely hampered due to the mismatch between training and testing distributions. Since the target domain usually lacks labeled data, and domain shifts exist at acoustic and linguistic levels, it is challenging to perform unsupervised domain adaptation (UDA) for ASR. Previous work has shown that self-supervised learning (SSL) or pseudo-labeling (PL) is effective in UDA by exploiting the self-supervisions of unlabeled data. However, these self-supervisions also face performance degradation in mismatched domain distributions, which previous work fails to address. This work presents a systematic UDA framework to fully utilize the unlabeled data with self-supervision in the pre-training and fine-tuning paradigm. On the one hand, we apply continued pre-training and data replay techniques to mitigate the domain mismatch of the SSL pre-trained model. On the other hand, we propose a domain-adaptive fine-tuning approach based on the PL technique with three unique modifications: Firstly, we design a dual-branch PL method to decrease the sensitivity to the erroneous pseudo-labels; Secondly, we devise an uncertainty-aware confidence filtering strategy to improve pseudo-label correctness; Thirdly, we introduce a two-step PL approach to incorporate target domain linguistic knowledge, thus generating more accurate target domain pseudo-labels. Experimental results on various cross-domain scenarios demonstrate that the proposed approach effectively boosts the cross-domain performance and significantly outperforms previous approaches.
\end{abstract}


%
\IEEEpeerreviewmaketitle

\section{Introduction}
\IEEEPARstart{T}{he} performance of end-to-end (E2E) automatic speech recognition (ASR) systems has improved dramatically over the past years \cite{li2021recent,cheng2022eteh,huang2019exploring,miao2020online,miao2020transformer} due to the advanced neural network architectures, improved training criteria, and large amounts of training data. However, the performance degradation on the cross-domain data is still a challenging issue for ASR due to the domain shift between the training and testing data. Since it is impossible to cover all test domains in the training data, applying domain adaptation \cite{bell2020adaptation} for a new target domain is of great interest in the application of ASR.

Domain adaptation aims to transfer a model trained on the source data to a given target domain in supervised or unsupervised conditions.
When the labeled data is available in the target domain, the supervised domain adaptation is straightforward since we could simply use the labeled data to fine-tune the source model \cite{sim2018domain,hou2021exploiting}. 
However, since the labeled target data is costly to collect, the unsupervised domain adaptation (UDA) scenario, where no labeled data is available in the target domain, is more desired in real-world applications.

Existing UDA approaches tackle the lack of labeled data issue from different aspects.
An intuitive solution is to synthesize target domain data \cite{li2017large,hsu2017unsupervised,hosseini2018multi,li2020developing,baskar2021eat,yue2022exploring}.
However, these approaches require a specific design for the target domain \cite{li2017large} or careful tuning of the data generation model \cite{hosseini2018multi}, making them inconvenient when extending to an arbitrary new domain or large-scale applications.
Another category is domain-invariant feature learning with distribution matching approaches \cite{hou2021cross,sun2018domain}, which aims to learn a domain-invariant representation while being class-discriminative on the source domain.
However, the class-discriminative representations on the target domain cannot be easily guaranteed.
Thus, it would fail under certain domain mismatch scenarios \cite{li2020rethinking,zhao2019learning,liu2019transferable}.

Recently, self-supervised learning (SSL) based pre-training~\cite{devlin2019bert,hwang2021large,misra2021comparison,hsu2021robust} and pseudo-labeling (PL)~\cite{khurana2021unsupervised,higuchi2021momentum} are shown to be effective for UDA of the E2E-ASR model by directly training on the unlabeled target data with self-supervision.
On the one hand, both SSL and PL approaches are simple in practice and do not need customization for specific domains.
Thus, they are convenient to be applied to any new domain and large-scale applications.
On the other hand, SSL and PL are shown to be robust under various domain mismatch conditions \cite{hsu2021robust,khurana2021unsupervised,higuchi2021momentum}.

Nonetheless, existing literature typically focused on one aspect to address the UDA problem, e.g., with SSL~\cite{hsu2021robust}, online~\cite{higuchi2021momentum} or offline PL~\cite{khurana2021unsupervised}. This practice failed to realize the full potential of self-supervision for UDA. To push the limits of UDA, in this work, we first identify the weaknesses of SSL and PL when applying them to the UDA scenario and propose innovative solutions to address them. Then, we seamlessly integrate the improved SSL and PL approaches into a systematic UDA framework to boost \textbf{C}ross-dom\textbf{A}in \textbf{S}peech recogni\textbf{T}ion with Se\textbf{L}f-Sup\textbf{E}rvision, namely, \textbf{\method}. 

In summary, the major novelties of this paper are as follows:

\begin{itemize}

\item \emph{Dual-Branch PL (DPL):} There are two vital challenges in existing online PL approaches: Firstly, since the self-generated pseudo-labels are used as the supervision, the errors would be accumulated and cause the error accumulation~\cite{jiang2020implicit} (or the confirmation bias~\cite{arazo2020pseudo}) issue, degrading the performance and sometimes driving the training to collapse. Secondly, due to the lack of confidence filtering design for online PL, all pseudo-labels are used in training, including the extremely noisy pseudo-labels. To address these challenges, on the one hand, DPL proposes to break the error accumulation chain by using an auxiliary branch to generate pseudo-labels. On the other hand, DPL utilizes a specifically designed confidence estimation that discards the CTC blank scores, thus robustly improving the online PL. Extensive experiments demonstrated the advantage of DPL over existing online PL approaches.
\item \emph{Uncertainty-Aware Confidence Filtering (UCF):} Most existing filtering methods for offline PL utilize decoding scores as confidence estimation to rule out the noisy pseudo-labels~\cite{kahn2020self,park2020improved}. However, this confidence estimation is unreliable when ASR networks are poorly calibrated~\cite{li2021confidence}. UCF addressed this issue by adaptively utilizing uncertainty~\cite{gal2016dropout} and confidence estimations to select pseudo-labels in offline PL, where the combination hyper-parameters in UCF are adaptively determined on the development set. Experiments demonstrated that UCF can outperform existing filtering approaches without tuning hyper-parameters.
\item \emph{Two-Step PL:} Two-step PL is an empirically motivated approach that utilizes the offline PL to refine the online PL linguistically. Although a simple combination of two types of PL approaches, it consistently outperforms either one of them in practice.
\item \emph{Continued Pre-Training with Data Replay:} Catastrophic forgetting is a critical issue in continual learning. Since continued pre-training~\cite{gururangan2020don,hsu2021robust} is proven effective and used in our approach, we examine whether catastrophic forgetting is severe here by evaluating the effectiveness of data replay~\cite{hu2022how}, which is widely adopted to address knowledge forgetting. Based on experimental findings, we provide suggestions on continued pre-training strategies given different fine-tuning strategies.
\end{itemize}

We performed detailed experiments on various cross-domain datasets and showed that the proposed approach \method could effectively boost the cross-domain performance of ASR and consistently outperforms previous UDA approaches. 

The rest of the paper is organized as follows. In \autoref{sec:releted} we formulate the UDA problem and review related works. Then the proposed approach \method is introduced in \autoref{sec:proposed}. We describe experimental settings in \autoref{sec:experiment}, and then present experimental results in \autoref{sec:results}. Finally, \autoref{sec:conclusion} concludes the paper.

\section{Preliminaries and Related Work}
\label{sec:releted}
We first define the notations. In an ASR model,
the acoustic feature $\mathbf{X}$ is first processed by a feature transformation function $g: \mathcal{X} \mapsto \mathcal{Z}$ to generate the latent feature representation $\mathbf{Z}$. Then the projection function $h: \mathcal{Z} \mapsto \mathcal{P}$ gives the final prediction of the ASR network $\mathbf{P}$. Finally, the decoding function $d: \mathcal{P} \mapsto \mathcal{Y}$ generates the transcription $\mathbf{Y}$. The composite transformation of the ASR network is $f=g \circ h: \mathcal{X} \mapsto \mathcal{P}$. And $f$, $g$, $h$ are parameterized by $\theta$, $\phi$, $\psi$ respectively.

Then we formulate the UDA problem. Suppose there are two different domains: source domain $S$ and target domain $T$.
In source domain, there is a large unlabeled dataset $\mathbb{U}^{S}=\left\{\mathbf{X}_{1}^{S}, \ldots, \mathbf{X}_{N}^{S}\right\}$ and a small labeled subset $\mathbb{L}^{S}=\left\{\left(\mathbf{X}_{1}^{S}, \boldsymbol{Y}_{1}^{S}\right), \ldots,\left(\mathbf{X}_{M}^{S}, \boldsymbol{Y}_{M}^{S}\right)\right\}$, where $M \leq N$ and $\left(\mathbf{X}, \boldsymbol{Y}\right)$ denote an feature-transcription pair. In target domain, only an unlabeled dataset $\mathbb{U}^{T}=\left\{\mathbf{X}_{1}^{T}, \ldots, \mathbf{X}_{O}^{T}\right\}$ is available. The goal of UDA is to improve the ASR performance on the target domain using all above datasets. In this work, we also assume the target domain style text corpus is available to train a target domain LM, as it is easy to collect. Moreover, a development set, i.e., a small labeled source domain dataset, is used during the training process.

The major challenge of the UDA scenario is the lack of labeled target data so that we can't directly perform supervised training on the target domain. In the following, we introduce mainstream UDA approaches for ASR, which utilize unlabeled target data in different ways. Briefly, domain-invariant feature learning uses a distribution match loss to match the distribution between labeled and unlabeled data. SSL-based approach exploits a self-supervised loss to learn a better representation for target domain distribution. PL-based approaches generate pseudo-labels for unlabeled data and then train the model on it.

\subsection{Domain-Invariant Feature Learning}
Learning domain-invariant features with distribution matching is a prominent UDA approach in many areas, which aims to generalize better to the target domain by minimizing the differences between the intermediate feature representations of the source and target domain. The distribution matching loss could be formulated as $D\left(g_{\phi}(X^{S}) \| g_{\phi}(X^{T})\right)$, where $D$ is the distance metric.
Specifically, some works \cite{hou2021cross} used Maximum Mean Discrepancy to measure and reduce the differences between two domains. Others \cite{sun2018domain,anoop2021unsupervised} chose domain adversarial training \cite{ganin2015unsupervised} to learn a domain-invariant representation to fool the domain classifier, which can be viewed as minimizing the Jensen-Shannon (JS) divergence \cite{ganin2016domain}. However, such practice is unreliable when label distribution shift~\cite{li2020rethinking}, conditional distribution shift~\cite{zhao2019learning}, or large domain discrepancy \cite{liu2019transferable} exist.

\subsection{Self-Supervised Learning}
Since SSL-based pre-training is shown to be effective in exploiting unlabeled data for ASR~\cite{baevski2020wav2vec,pascual2019learning,chung2019unsupervised,liu2021tera,hsu2021hubert,yang2021superb,chung2021w2v,gao2021pre,gao2022self,gao2021data}, it is suitable to use SSL to utilize unlabeled source and target data for UDA~\cite{hsu2021robust}. 

Robust wav2vec 2.0 \cite{hsu2021robust} showed that joint pre-training on unlabeled target data could improve target domain performance when fine-tuning on the labeled source data. However, since joint pre-training is time-consuming~\cite{hsu2021robust}, thus continued pre-training~\cite{gururangan2020don} would be more appealing with low computation budget. Nonetheless, continued pre-training may suffer from the catastrophic forgetting of source knowledge~\cite{hu2022how}, which we study in detail in this work. 

Joint supervised and self-supervised training \cite{bai2022joint,talnikar2021joint,hwang22c_interspeech} was shown to be better than self-supervised pre-training in settings with domain mismatch. However, in the studied UDA scenario, we failed to achieve better performance than the pure self-supervised approach with such joint training method, specifically, applying wav2vec 2.0 loss on unlabeled target data and CTC loss on labeled source data. 
One possible reason is that wav2vec 2.0 and CTC encourage different representation patterns~\cite{pasad2021layer}. Minimizing two losses alternatively with separate optimizers \cite{talnikar2021joint}, or using some separate layers for each loss \cite{hwang22c_interspeech} can possibly address this issue.
We stick to the self-supervised pre-training as it is more easily implemented and leave these explorations in the future.

\subsection{Pseudo-labeling}
Pseudo-labeling (PL)~\cite{sohn2020fixmatch,zhang2021flexmatch,chen2022debiased}, also known as self-training, is widely used for semi-supervised learning due to its effectiveness and simplicity.

\begin{figure}[!tbp]
    \centering
	\subfloat[Offline PL] 
	{ \label{fig:offline_PL}
		\includegraphics[width=0.65\linewidth]{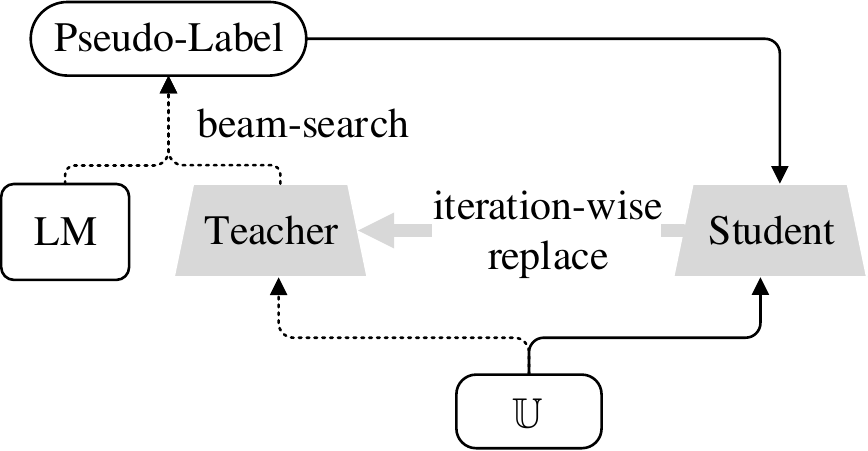}
	}

	\subfloat[Online PL] 
	{ \label{fig:online_PL}
		\includegraphics[width=0.55\linewidth]{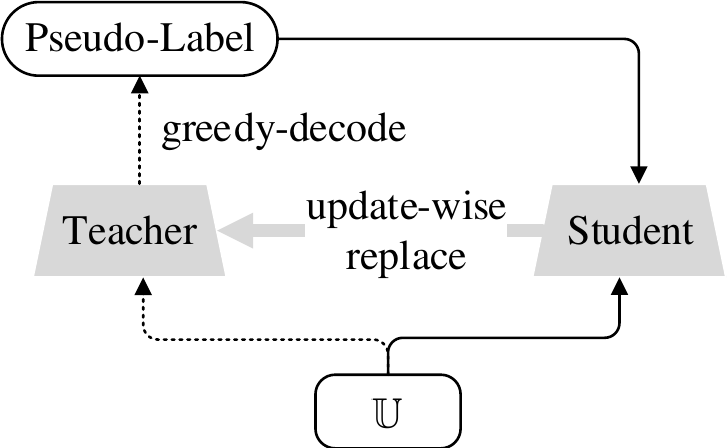}
	}
	\caption{Illustration of generating pseudo-labels and updating the teacher model in offline and online PL. The dashed line denotes inference mode and the solid line denotes training mode\protect\footnotemark[1]. } 
	\label{fig:comparison_PL}
\end{figure}

\footnotetext[1]{Dropout and data augmentations are enabled in training mode but disabled in inference mode.}  
\setcounter{footnote}{1}

There are two mainstream implementations of PL in ASR: offline PL~\cite{weninger2020semi,masumura2020sequence,li2019semi,synnaeve2019end,xu2020iterative} and online PL~\cite{chen2020semi,manohar2021kaizen}. 
Both approaches use a teacher to generate pseudo-labels and a student to utilize pseudo-labels. The student is trained with back-propagation while the teacher is replaced with the student after certain intervals, which means the teacher and student have the same structure.

Online and offline PL differ in generating pseudo-labels and updating the teacher model. We illustrate the differences in \autoref{fig:comparison_PL}.
In online PL, pseudo-labels are generated online in each training update. In offline PL, pseudo-labels are generated offline before each training iteration\footnote{Each training iteration consists of multiple training updates.}. When we formulate online/offline PL in the teacher-student framework, the fundamental difference between online and offline PL is the update frequency of the teacher model: online and offline PL update the teacher model after each training update and each training iteration, respectively.

Conventionally, offline PL is used as an LM-based PL approach where the LM is integrated when generating pseudo-labels~\cite{park2020improved,xu2020iterative,liao2013large,lanchantin2015development,lanchantin2016selection}. On the contrary, Since the online generation of pseudo-labels requires the decoding speed to be fast, online PL approaches usually apply greedy-decoding, thus being an LM-free PL approach. Despite its simplicity, online PL could achieve competitive performance compared with offline PL \cite{manohar2021kaizen}.

Since pseudo-labels are noisy, filtering techniques could be applied to select pseudo-labels with better quality. Confidence filtering approaches~\cite{kahn2020self,park2020improved} utilize the decoding score to select pseudo-labels. However, the poorly calibrated neural networks could produce over-confident erroneous predictions and make the confidence filtering unreliable \cite{li2021confidence}. To improve the confidence estimation, some model-based approaches utilize an additional confidence estimation module \cite{li2021confidence}. But they also complicate the training procedure. 
As an alternative, some work~\cite{khurana2021unsupervised} used the uncertainty estimation of the ASR model to filter out the erroneous pseudo-labels. The uncertainty could be modeled with the prediction variance~\cite{zheng2021rectifying} and estimated via the Monte Carlo dropout~\cite{gal2016dropout}.
Apart from the filtering techniques, prediction combination~\cite{berthelot2019mixmatch,chen21c_interspeech} can also be used to improve the quality of pseudo-labels.

In terms of the UDA scenario for ASR, offline PL is shown to be effective in \cite{khurana2021unsupervised,hwang2021large}, where \cite{khurana2021unsupervised} applied the dropout-based uncertainty filtering and \cite{hwang2021large} utilized the model-based confidence filtering~\cite{li2021confidence}.
Online PL was explored in MPL~\cite{higuchi2021momentum}, which utilizes the EMA technique to stabilize training.

\section{Proposed Approach}
\label{sec:proposed}

\begin{algorithm}[!t]
	\caption{CASTLE algorithm.}
	\begin{small}
	
	{\bf Input:}
	Labeled source dataset $\mathbb{L}^{S}$, unlabeled target dataset $\mathbb{U}^{T}$, SSL model $\mathcal{M}_\text{SSL}$ pre-trained on $\mathbb{U}^{S}$, target domain LM $\mathcal{M}_\text{LM}$, hyper-parameters $\alpha$, $K$, $c_\text{on}$. \\
	{\bf Output:} 
	ASR model $\mathcal{M}_\text{ASR}$.
	\begin{algorithmic}[1]
	
		\State Continued pre-train $\mathcal{M}_\text{SSL}$ on $\mathbb{U}^{T}$;
		\State Add random initialized linear layer on $\mathcal{M}_\text{SSL}$ to produce $\mathcal{M}_\text{ASR}$;
		\Repeat
		\State Draw batches $\mathbf{B}^{S}$, $\mathbf{B}^T$ of the same size from $\mathbb{L}^{S}$ and $\mathbb{U}^{T}$;
		\State Filter $\mathbf{B}^T$ with the confidence score to produce $\mathbf{B}^{T\prime}$;
		\State Compute DPL loss $\mathcal{L}_{\text{DPL}}$ with $\mathbf{B}^{S}$ and $\mathbf{B}^{T\prime}$;
		\State Update the model $\mathcal{M}_\text{ASR}$ with $\mathcal{L}_{\text{DPL}}$;
		\Until{maximum updates for online PL are reached}
		\Repeat
		\State  Decode $\mathbb{U}^{T}$ with $\mathcal{M}_\text{ASR}$ and $\mathcal{M}_\text{LM}$ to get $\mathbb{\hat{L}}^{T}$;
		\State Filter $\mathbb{\hat{L}}^{T}$ with UCF to produce $\mathbb{\hat{L}}^{T\prime}$;
		\Repeat
		\State Draw a batch $\mathbf{\hat{B}}^{T}$ from $\mathbb{\hat{L}}^{T\prime}$;
		\State Compute offline PL loss $\mathcal{L}_{\text{offlinePL}}$ with $\mathbf{\hat{B}}^{T}$;
		\State Update the model $\mathcal{M}_\text{ASR}$ with $\mathcal{L}_{\text{offlinePL}}$;
		\Until{maximum updates for current iteration are reached}
		\Until{maximum iterations for offline PL are reached}
	\end{algorithmic}
	\end{small}
	\label{alg:castle}
\end{algorithm}

The proposed CASTLE approach consists of the following stages. Given a pre-trained SSL model, e.g., wav2vec 2.0, we first alleviate the pre-training mismatch by continued pre-training on the unlabeled target domain $\mathbb{U}^{T}$. After that, we perform the domain-adaptive fine-tuning to resolve the fine-tuning mismatch. Specifically, we utilize the two-step PL to conduct the online and offline PL step by step, where the dual-branch PL is used as the online PL approach and the uncertainty-aware confidence filtering is used during offline PL. 
In this work, we concentrate on CTC~\cite{graves2006connectionist} models, which enable non-autoregressive decoding and achieve state-of-the-art results for many low-resource scenarios~\cite{baevski2020wav2vec,zhang2020pushing,likhomanenko2020slimipl,lugosch2022pseudo}. We summarize the complete training procedure in \autoref{alg:castle}. And the details are described as follows.

\subsection{Continued Pre-Training with Data Replay}

\begin{figure}[!t]
	\centering
	\includegraphics[width=0.6\columnwidth]{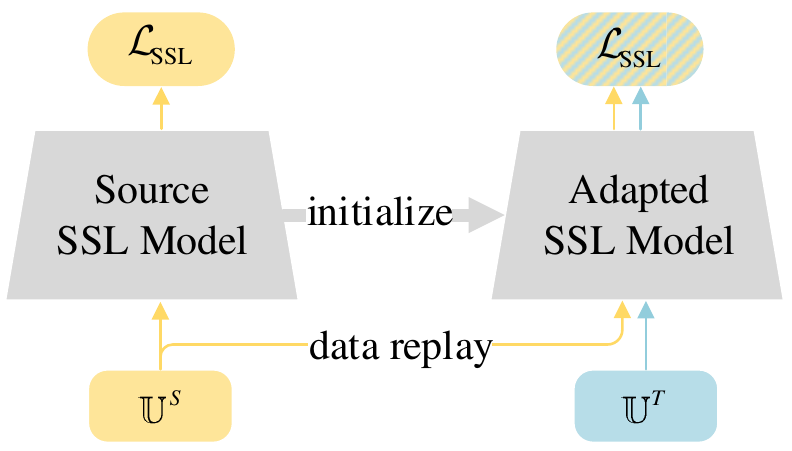}
	\caption{Illustration of continued pre-training with data replay.} 
	\label{fig:CPT-DR}
\end{figure}

Continued pre-training on the target domain could effectively alleviate the domain mismatch of the SSL pre-training model while remaining light-weight computation \cite{hsu2021robust}. However, the continued pre-training could lead to the catastrophic forgetting of the source knowledge because the model only observes the target data during this stage. Suppose the final SSL model is only fine-tuned on the labeled source data, and we expect generalization to the target domain. In that case, it is preferable that both source and target knowledge are well aligned in the SSL model. 

To resolve this issue, as shown in \autoref{fig:CPT-DR}, we could replay the source data during continued pre-training. Specifically, we select source and target data with a ratio $p_{s/t}$. The ratio $p_{s/t}$ is smaller than 1 since the source model is already well trained on the source data. And the source data is only used as the regularization to avoid knowledge forgetting.

\subsection{Online PL with Dual-Branch PL}
\label{sec:DPL}

In the online PL approach, since pseudo-labels are generated in each update, the error accumulation issue is much more severe. Dual-branch PL (DPL) could effectively alleviate this issue. We explain the details of DPL as follows.

\subsubsection{Dual-Branch Learning}

\begin{figure}[t!]
	\centering
	\includegraphics[width=0.7\columnwidth]{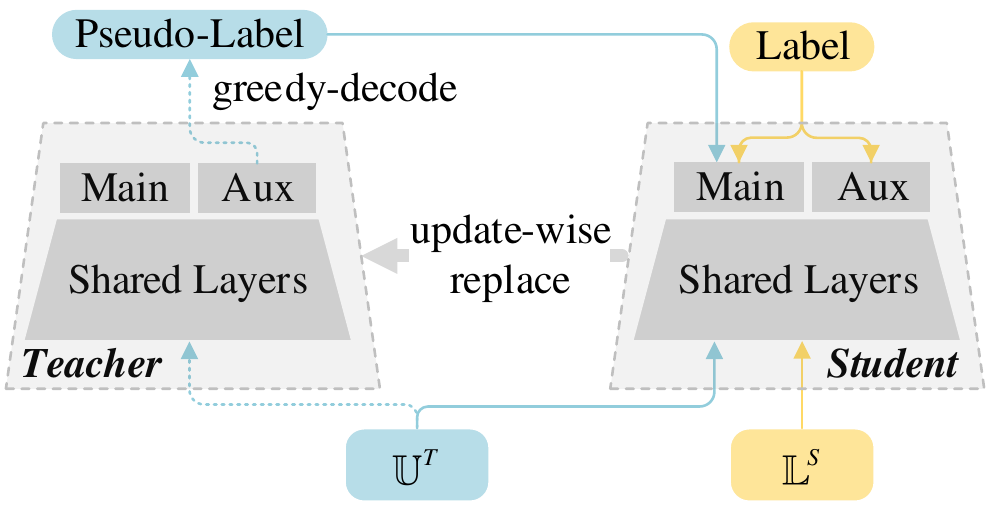}
	\caption{Illustration of dual-branch learning in DPL.} 
	\label{fig:dual_branch}
\end{figure}

The main idea of DPL is the dual-branch learning, which utilizes partially separated parameters to generate and exploit pseudo-labels, thus breaking the error accumulation chain. We illustrate the training procedure of dual-branch learning in \autoref{fig:dual_branch} and explain it as follows.

In dual-branch learning, the model consists of a shared feature transformation function $g_{\phi}$, and two projection functions $h_{\psi_{a}}$ and $h_{\psi_{m}}$, where $h_{\psi_{a}}$ is the auxiliary branch and $h_{\psi_{m}}$ is the main branch. These two branches are stacked on top of $g_{\phi}$. Note that the projection function $h$ corresponds to the final linear layer of the CTC model in this work.

For samples in the unlabeled target dataset $\mathbb{U}^{T}$, the model generates pseudo-labels from the auxiliary branch:
\begin{equation}
\hat{\mathbf{Y}}=\operatorname{greedy-decode}\left(\widetilde{h}_{\psi_{a}}(\widetilde{g}_{\phi}(\mathbf{X}))\right), \mathbf{X} \in \mathbb{U}^{T},
\end{equation}
where $\widetilde{h}$ and $\widetilde{g}$ are $h$ and $g$ in the inference mode.

Then the online PL loss is computed between predictions from the main branch and the above pseudo-labels:
\begin{equation}
\label{equ:loss_online_PL}
\mathcal{L}_{\text{onlinePL}}=\operatorname{CTC}\left(h_{\psi_{m}}(g_{\phi}(\mathbf{X}^{\prime})),\hat{\mathbf{Y}}\right), \mathbf{X} \in \mathbb{U}^{T},
\end{equation}
where $\mathbf{X}^{\prime}$ is the augmented version of $\mathbf{X}$.

For samples in the labeled source dataset $\mathbb{L}^{S}$, both the auxiliary branch and the main branch are trained on them as:
\begin{equation}
\begin{aligned}
\mathcal{L}_{\text{SUP}}=&\operatorname{CTC}\left(h_{\psi_{a}}(g_{\phi}(\mathbf{X}^{\prime})) ,\mathbf{Y}\right) \\ &+ \operatorname{CTC}\left(h_{\psi_{m}}(g_{\phi}(\mathbf{X}^{\prime})) ,\mathbf{Y}\right), (\mathbf{X}, \mathbf{Y}) \in \mathbb{L}^{S},
\end{aligned}
\end{equation}

Finally, the total loss is:
\begin{equation}
\label{equ:DPL}
\mathcal{L}_{\text{DPL}} = \mathcal{L}_{\text{SUP}} + \alpha \mathcal{L}_{\text{onlinePL}}, 
\end{equation}
where $\alpha$ is a hyper-parameter to be tuned.

In dual-branch learning, both labeled source data and unlabeled target data is used to optimize the main branch, while only labeled source data is used to optimize the auxiliary branch. Since pseudo-labels are generated with the auxiliary branch while exploited by the main branch, DPL could break the error accumulation chain. And since the shared feature transformation function is optimized by both source and target data, the latent feature representation could generalize to both domains. Thus the auxiliary branch that is only optimized with source data could also produce accurate enough pseudo-labels on the target domain. Moreover, the auxiliary branch will be discarded after training, thus keeping the same inference computation.

\subsubsection{One Stage Training with Confidence Filtering}
\label{sec:DPL_filter}

Previous online PL approaches usually consist of supervised training as the first stage and PL training as the second stage. This practice could avoid training on noisy pseudo-labels in the beginning but neglects the fact that pseudo-labels could still be noisy in the second stage. This issue is especially serious with domain mismatches.

To resolve this issue, we use the confidence filtering to automatically choose the pseudo-labels since the confidence score could roughly reflect the quality of pseudo-labels.

Although there are many approaches to estimate the confidence score for offline PL~\cite{kahn2020self,park2020improved}, few attempts have been made for online PL. In offline PL, pseudo-labels are mostly generated with beam-search, which naturally generates a decoding score that can be used as the confidence estimation. On the other hand, greedy search is used in online PL, and the decoding score for greedy search is not properly defined.

To fill this gap, we design a confidence estimation for online PL. Specifically, we compute the confidence score by taking the mean value of the maximum scores in all frames, which shares a similar idea with the decoding score in offline PL. Our unique design is: to avoid the trivial solution that the model always predicts blank tokens in CTC~\cite{likhomanenko2020slimipl,manohar2021kaizen}, we discard the frames where the maximum score belongs to the blank token. A similar idea is used in \cite{chen2017confidence} to improve the confidence measure of the CTC-based ASR model.

Formally, given an sample $\mathbf{X} \in \mathbb{U}^{T}$, the prediction of the ASR network is computed as:
\begin{equation}
\mathbf{P}= \operatorname{log-softmax}\left(f_{\theta}\left(\mathbf{X}\right)\right), \mathbf{X} \in \mathbb{U}^{T},
\end{equation}
where $\mathbf{P} \in \mathbb{R}^{T \times V}$. $T$ is the number of frames and $V$ is the output vocabulary size including the blank token.

Then we remove the frames $\mathbf{P_t}$ where the maximum score belongs to the blank token. And the resulting prediction is $\mathbf{P^{\prime}} \in \mathbb{R}^{T^{\prime} \times V}$.

Finally, the confidence score is estimated as: 
\begin{equation}
\label{equ:confidence}
\begin{gathered}
C_\text{on}=\operatorname{exp}\left(\frac{1}{T^{\prime}}\sum_{t=1}^{T^{\prime}}  \max_{v}\left(\mathbf{P^{\prime}}_{t,v}\right)\right)
\end{gathered}
\end{equation}

It is worth noting that this confidence estimation does not require extra parameters or computations, thus fitting in the online PL approach.

To this end, the PL loss $\mathcal{L}_{\text{onlinePL}}$ will only include samples with confidence $C_\text{on} \geq c_\text{on}$, where $c_\text{on}$ is the filtering threshold.

\subsubsection{Sampling Strategy}

We utilize a fixed sampling strategy where the labeled source data and unlabeled target data are sampled with the same batch size in each update. In this way, a stable supervised signal exists in each update to regularize the PL training. And we could control the weight of the unlabeled data with the PL loss weight $\alpha$ in \autoref{equ:DPL}. A proper $\alpha$ value could effectively stabilize training without using EMA, which is shown to be important for stabilization if we randomly sample from the mixed dataset of both labeled and unlabeled data~\cite{higuchi2021momentum} or only sample the unlabeled data~\cite{manohar2021kaizen}. Therefore, we could avoid maintaining an additional EMA version of the model, thus being more memory efficient. 

\subsection{Offline PL with Uncertainty-Aware Confidence Filtering}

Similar to online PL, offline PL generates pseudo-labels for the unlabeled target data with the teacher model as:
\begin{equation}
\label{equ:offline_PL_beam_search}
\hat{\mathbf{Y}}=\operatorname{beam-search}\left(\widetilde{f}_{\xi}(\mathbf{X}), \mathcal{M}_\text{LM}\right), \mathbf{X} \in \mathbb{U}^{T},
\end{equation}
where $\xi$ is the parameter of the teacher model, which is replaced with the up-to-date student model $\theta$ after each iteration of offline PL. $\widetilde{f}$ is $f$ in the inference mode. And $\mathcal{M}_\text{LM}$ is the target domain LM.

Then, the offline PL loss is computed as:
\begin{equation}
\label{equ:loss_offline_PL}
\mathcal{L}_{\text{offlinePL}}=\operatorname{CTC}\left(f_\theta(\mathbf{X}^{\prime}),\hat{\mathbf{Y}}\right), \mathbf{X} \in \mathbb{U}^{T},
\end{equation}
where $\theta$ is parameter of the student model and $\mathbf{X}^{\prime}$ is the augmented version of $\mathbf{X}$.

\begin{figure}[t!]
	\centering
	\includegraphics[width=0.8\columnwidth]{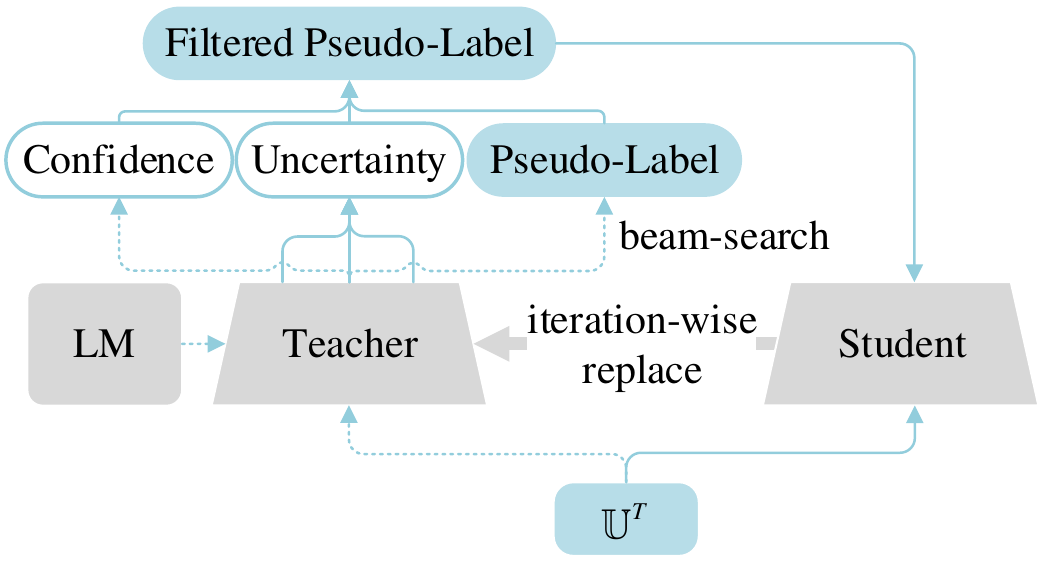}
	\caption{Illustration of offline PL with uncertainty-aware confidence filtering (UCF). Pseudo-labels and confidence scores are generated in the inference mode (dashed line). And the uncertainty estimation is generated in the training mode\protect\footnotemark[3] (solid line).} 
	\label{fig:UCF}
\end{figure}

\footnotetext[3]{Unlike previous notations, the training mode here enables dropout but disables data augmentation.}  
\setcounter{footnote}{3}

To improve the quality of offline pseudo labels $\hat{\mathbf{Y}}$, we could adopt filtering approaches based on confidence or uncertainty estimations. However, the confidence score is unreliable for a poorly calibrated neural network and we find low uncertainty alone also can't guarantee the accuracy of selected pseudo-labels. Intuitively, pseudo-labels with higher confidence and low uncertainty are more likely to be correct. Therefore, we integrate the confidence and uncertainty in one framework for filtering, i.e., uncertainty-aware confidence filtering (UCF). We illustrate UCF in \autoref{fig:UCF} and describe the details as follows.

The decoding score from the beam-search decoding in \autoref{equ:offline_PL_beam_search} could be used as the confidence score $C_\text{off}$. And the $C_\text{off}$ is normalized by the sample length.

As for the uncertainty, we adopt Monte Carlo dropout~\cite{gal2016dropout} to estimate the prediction variance with some modifications for the ASR task. The procedure is described below.

Firstly, we compute pseudo-labels with dropout enabled for $K$ times:
\begin{equation}
\hat{\mathbf{Y}}_{k}=\operatorname{beam-search}\left(f_{\xi}(\mathbf{X}), \mathcal{M}_\text{LM}\right), \mathbf{X} \in \mathbb{U}^{T},
\end{equation}

Then, the uncertainty could be estimated with the prediction variance as:
\begin{equation}
U = \frac{1}{K |\hat{\mathbf{Y}}|} \sum_{k=1}^{K} D\left( \hat{\mathbf{Y}}_{k}\| E(\hat{\mathbf{Y}}_{k})\right)
\end{equation}
where $D$ is the distance metric. Since $\mathbf{Y}$ is the text, we utilize the edit distance (Levenshtein distance) here.

\begin{figure*}[t!]
	\centering
	\includegraphics[width=1.5\columnwidth]{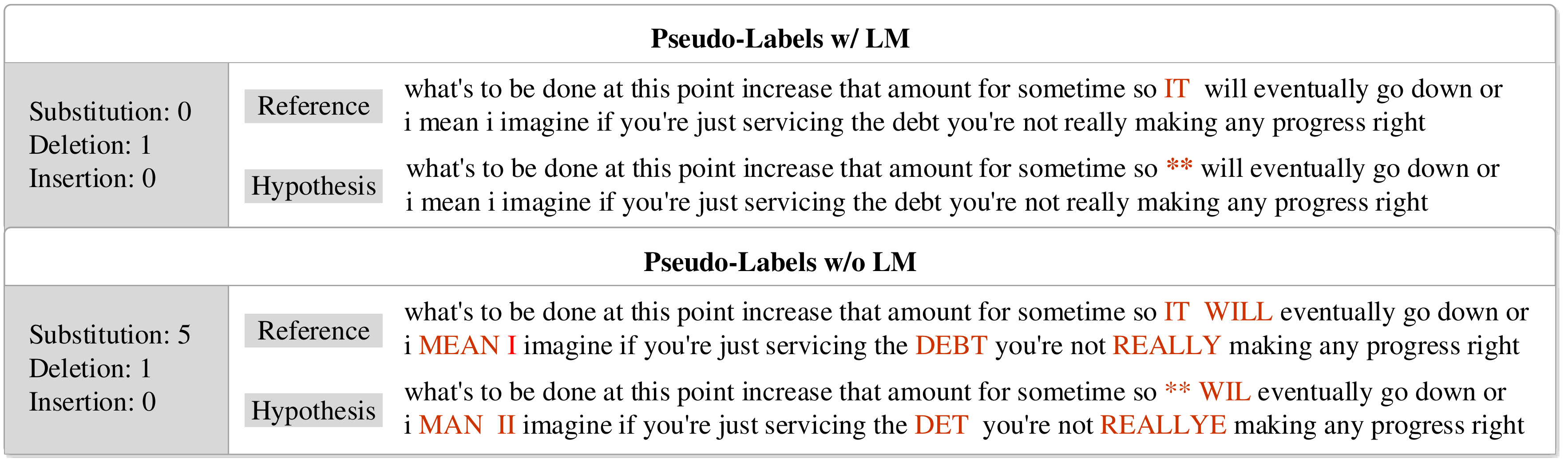}
	\caption{Illustration of errors in pseudo-labels generated w/ or w/o LM. The same CTC-based ASR model is used to generate these pseudo-labels.} 
	\label{fig:errors}
\end{figure*}

However, it is unknown how to compute the mean text $E(\hat{\mathbf{Y}})$. Since the decoding results with dropout could be viewed as the deviations of the one without dropout, we could use the approximation that $\hat{\mathbf{Y}} = E(\hat{\mathbf{Y}_{k}})$, which means using the decoding result without dropout as the mean text.

To this end, the uncertainty could be estimate with:
\begin{equation}
U = \frac{1}{K |\hat{\mathbf{Y}}|} \sum_{k=1}^{K} \operatorname{edit-distance}\left( \hat{\mathbf{Y}}_{k} \| \hat{\mathbf{Y}}\right) 
\end{equation}

Compared with the uncertainty estimation in previous work \cite{khurana2021unsupervised}, our estimation uses the mean of edit distances instead of the maximum since it is more mathematically intuitive.

Moreover, we find the model tends to select shorter samples in some cases. But those selected shorter samples are still erroneous. Therefore, we add a length bonus term $log(|\hat{\mathbf{Y}}|)$ to alleviate this issue:

Finally, the UCF strategy would select samples with:
\begin{equation}
C_\text{off} - \gamma U + \eta \log(|\hat{\mathbf{Y}}|) >= c_\text{off}
\end{equation}

where $c_\text{off}$ is the filtering threshold. $\gamma$ and $\eta$ are hyper-parameters that can be directly estimated on the development set, thus alleviating the effort for hyper-parameter tuning.

The estimation of the uncertainty requires multiple forward computations for each sample. For offline PL, since the estimation is only conducted after each iteration, the increased computation is tolerable. But it will significantly increase the training time for online PL since the estimation is conducted in each update. Therefore, we only apply UCF for offline PL.

\subsection{Two-Step PL}

Benefiting from the continuous improvement of pseudo-labels~\cite{manohar2021kaizen}, the ASR model could be fast adapted to the target domain with online PL. 
However, online PL purely relies on the source domain transcripts to learn linguistic knowledge, while the target domain linguistics are neglected. Consequently, pseudo-labels generated in online PL are likely to be incorrect in linguistics. 
We illustrate this effect in \autoref{fig:errors}. After the adaptation with online PL, we use the adapted model to generate pseudo-labels with or without LM. We take an utterance from the SWBD training set as an example. The ground-truth labels are denoted as the reference and pseudo-labels are denoted as the hypothesis. When LM is used, there is only one deletion error. However, when decoding without LM, there are five additional substitution errors, roughly correct in acoustic but incorrect in linguistics. 

Training on these erroneous pseudo-labels would be sub-optimal. Thus, we could consider using the target domain LM to improve the accuracy of pseudo-labels. Although there are no transcripts for the target domain samples, it is not that hard to collect some text with the target domain style. 
Target domain LM could be naturally integrated into offline PL by generating pseudo-labels with the beam-search decoding and LM. If the target domain LM is used, offline PL could inject the linguistic knowledge of the target domain into the ASR model by training on these pseudo-labels.
To this end, we propose the two-step PL strategy that can be summarized as: online PL for on-the-fly adaptation followed by offline PL for further linguistic refinement. In comparison to pure online PL, two-step PL addresses linguistically incorrect pseudo-labels with the refinement of offline PL that relies on LM-based decoding. Compared with pure offline PL, two-step PL utilizes online PL to offer a superior seed model for offline PL, thereby improving final performance.

\section{Experimental Setup}
\label{sec:experiment}
\subsection{Corpus}

\subsubsection{Source Domain Corpus}
The source domain pre-trained model is the wav2vec 2.0 base model pre-trained on Librispeech 960h\footnote{The pre-trained model is downloaded from the wav2vec 2.0 repository: \href{https://github.com/pytorch/fairseq/blob/master/examples/wav2vec}{https://github.com/pytorch/fairseq/blob/master/examples/wav2vec}}. Consequently, LibriSpeech~\cite{panayotov2015librispeech} is used as the source domain dataset. We use the entire 960h training set as the unlabeled source domain dataset and the 100h subset as the labeled source domain dataset. 

\subsubsection{Target Domain Corpus}
We use several cross-domain corpora to evaluate the performance under different mismatched situations.
Specifically, we use Indian English Common Voice (CV), TED-LIUM v3 (TED) \cite{hernandez2018ted} and SwitchBoard (SWBD) \cite{godfrey1993switchboard} as unlabeled target domain datasets. Indian English Common Voice (CV) is extracted from the original Common Voice corpus \cite{ardila2020common} by selecting Indian accented data. We split the training, development and testing sets with the proportion of $8: 1: 1$, respectively.
For TED, standard train/dev/test splits are used. In terms of SWBD, the development set is RT-03S \cite{rt03} and the testing sets are Hub05 Eval2000 \cite{eval2000} SwitchBoard (H-SB) and CallHome (H-CH) subsets. All audios are re-sampled to 16kHz and transcripts are pre-processed to upper-case letters and no punctuation except apostrophes. Therefore, the corpora in source and target domain have the same speech and transcription formats. The details of the target domain datasets are shown in \autoref{tab:datasets}. 

\begin{table}[htbp]
  \centering
  \caption{The structure of target domain corpus (hours).}
  \label{tab:datasets}%
    \begin{tabular}{ccccccc}
    \toprule
    Split & CV & TED & SWBD \\
    \midrule
    Train & 48 & 452 & 319 \\
    Dev & 6 & 2 & 6 \\
    Test & 6 & 3 & 4 \\
    \bottomrule
    \end{tabular}%

\end{table}%

\subsubsection{Corpus for Target Domain LM}
Since the offline PL always requires decoding the training set with the target domain LM, the text corpus for LM should not contain the transcripts of the training set. For CV, we use the LM of Librispeech as the target domain LM since they are both read speech. For TED, we use the official LM training corpus of TED \cite{rousseau2014enhancing}, which is extracted from publicly available corpora distributed within the WMT 2013 machine translation evaluation campaign and is not overlapped with the training transcripts of TED. In terms of SWBD, we use transcripts of Fisher \cite{fisher} dataset since they are both conversational speech.

\begin{table*}[!t]
  \centering
  \caption{Comparison with unsupervised baselines and supervised toplines.}
  \label{tab:main}
  \resizebox{.85\textwidth}{!}{
    \begin{tabular}{lcccccccccccc}
    \toprule
    \multirow{3}[6]{*}{Method} & \multicolumn{2}{c}{Pre-Train Data} & \multicolumn{3}{c}{Fine-Tune Data} & \multicolumn{7}{c}{WER\%} \\ \cmidrule(lr){2-3} \cmidrule(lr){4-6}  \cmidrule(lr){7-13}        & Source & Target & Source & \multicolumn{2}{c}{Target} & \multicolumn{2}{c}{CV} & \multicolumn{2}{c}{TED} & \multicolumn{3}{c}{SWBD} \\
\cmidrule(lr){2-2} \cmidrule(lr){3-3} \cmidrule(lr){4-4} \cmidrule(lr){5-6} \cmidrule(lr){7-8} \cmidrule(lr){9-10} \cmidrule(lr){11-13}          & Unlabeled & Unlabeled & Labeled & Labeled & Unlabeled & Dev   & Test  & Dev   & Test  & RT03  & H-SB  & H-CH \\
    \midrule
    \midrule
    \multicolumn{13}{l}{\textit{\textbf{Unsupervised results}}} \\
    Wav2vec 2.0 & 960h  & ×     & 100h  & ×     & ×     & 35.7  & 36.2  & 10.2  & 10.6  & 35.9  & 25.8  & 34.1 \\
    \midrule
    Robust wav2vec & 960h  & all   & 100h  & ×     & ×     & 22.2  & 22.4  & 8.8   & 8.7   & 24.8  & 17.2  & 23.9 \\
     + DAT & 960h  & all   & 100h  & ×     & all    & 19.5  & 19.6  & 8.9   & 8.6   & 24.6  & 17.2  & 23.9  \\
     + MPL & 960h  & all   & 100h  & ×     & all   & 16.4  & 16.4  & 7.6   & 7.2   & 18.4  & 12.9  & 18.3 \\
     + DUST & 960h  & all   & 100h  & ×     & all   & 18.5  & 18.3  & 7.3   & 6.9   & 18.5  & 12.2  & 18.8 \\
    \midrule
    CASTLE (online PL) & 960h & all & 100h & ×   & all & \textbf{15.8} & \textbf{15.9} & 6.8 & 6.5 & 17.0 & 11.8 & 17.6 \\
    CASTLE (offline PL) & 960h & all & 100h & ×   & all & 17.7 & 17.6 & 7.2 & 6.9 & 17.3 & 11.2 & 17.6 \\
    CASTLE (two-step PL) & 960h & all & 100h & ×   & all & \textbf{15.8} & \textbf{15.9} & \textbf{6.6} & \textbf{6.3} & \textbf{16.1} & \textbf{10.8} & \textbf{17.1} \\
    \midrule
    \midrule
    \multicolumn{13}{l}{\textit{\textbf{Supervised results}}} \\
    Wav2vec 2.0 & 960h  & ×     & ×     & all   & ×     & 13.8  & 13.9  & 7.1   & 7.0   & 12.8  & 7.6   & 13.5 \\
    \midrule
    \multirow{1.5}[2]{*}{Robust wav2vec} & 960h  & all   & ×     & all   & ×     & 13.7  & 13.6  & 7.0   & 6.8   & 12.4  & \textbf{7.1} & 13.3 \\
          & 960h  & all   & 100h  & all   & ×     & \textbf{13.5} & \textbf{13.3} & \textbf{6.4} & \textbf{6.7} & \textbf{12.3} & 7.4   & \textbf{13.2} \\
    \bottomrule
    \end{tabular}%
    }
\end{table*}%

\subsection{Implementation Details}
All experiments are conducted using the fairseq \cite{ott2019fairseq} toolkit. The code is publicly available\footnote{ \href{https://github.com/zhu-han/CASTLE}{https://github.com/zhu-han/CASTLE}}.

For continued pre-training in \method and other compared approaches, we use the effective batch size of 89.6m samples. The total training updates are 20k. And the learning rate is decayed from $5\times10^{-5}$ without warmup. Continued pre-training applies the same time dimensional masking strategy with the source domain pre-training~\cite{baevski2020wav2vec}.

As for fine-tuning in \method and other compared approaches, the effective batch size is 51.2m samples. The maximum learning rate is $3\times10^{-5}$ and the tri-state learning rate schedule~\cite{baevski2020wav2vec} is adopted. The convolution feature encoder is fixed during fine-tuning. The masking strategy during fine-tuning is masking in both time and channel dimensions, similar to SpecAugment~\cite{park2019specaugment}.

In \method, the training updates for online PL is 20k. The offline PL consists of two iterations, where each iteration has 5k updates. There are various hyper-parameters in \method: DPL loss weight $\alpha$, online filtering threshold $c_\text{on}$, offline filtering threshold $c_\text{off}$, UCF's hyper-parameters $\gamma$, $\eta$ and $K$. Nonetheless, we find in practice that we can fix most hyper-parameters and automatically determine the others on the development set. Therefore, we do not need to tune them for a given dataset. Specifically, we set $\alpha = 1$, $c_\text{on} = 0.8$, $K=3$, $c_\text{off}$ is set to a value that $50\%$ of development set's pseudo-labels would be selected, $\gamma$ and $\eta$ are set by minimizing the WER of the selected $50\%$ of the development set's pseudo-labels. Note that fixing most or all hyper-parameters is widely adopted in existing approaches~\cite{sohn2020fixmatch,berthelot2019mixmatch,berthelot2021adamatch}.

As for evaluation, beam-search decoding with the 4-gram target domain LM is used to evaluate the performance. 

Note that the two-step PL approach is more complicated than the one-step PL approach. A simpler method would be more desirable in certain situations, such as large-scale ASR training. Hence, in addition to the two-step PL version of CASTLE, we introduce one-step PL versions of CASTLE that solely employ either online PL or offline PL. These three versions of CASTLE offer a trade-off between performance and complexity, allowing users to select a variation according to their preferences.

\section{Results}
\label{sec:results}

\subsection{Comparison with Previous Approaches}

We implement UDA approaches in previous literature and compare them with the proposed approach \method. Since most previous approaches tackle only pre-training or fine-tuning mismatch, we also implement their combinations to formulate stronger baselines that address both pre-training and fine-tuning mismatch. Specifically, we compare with:
\begin{itemize}
\item \emph{Wav2vec 2.0}: We directly fine-tune the source domain wav2vec 2.0 model with labeled source data for a maximum of 30k updates.
\item \emph{Robust Wav2vec}: We follow the efficient implementation in robust wav2vec 2.0~\cite{hsu2021robust}, i.e., continued pre-training on the target domain. Training updates for continued pre-training and fine-tuning are 20k and 30k respectively. Since the robust wav2vec 2.0 is consistently better than wav2vec 2.0, we use it as the pre-trained models in the following three approaches.
\item \emph{DAT}: Domain adversarial training (DAT)~\cite{sun2018domain} is used to learn domain-invariant features. The model is simultaneously optimized with the supervised loss on the source domain and the DAT loss between the source and target domain for 30k updates. The weight of the DAT loss is tuned on the development sets. 
\item \emph{MPL}: Momentum pseudo-labeling (MPL)~\cite{higuchi2021momentum} utilizes the EMA model to generate pseudo-labels in online PL, thus being more stable. The model is first trained on labeled source data for 10k updates, then trained with all data for another 20k updates. 
In the second stage, we randomly sample each batch from the mixed dataset of the source and target data as in \cite{higuchi2021momentum}. 
The discount factor of EMA is tuned on the development sets.
\item \emph{DUST}: Uncertainty-driven self-training (DUST)\cite{khurana2021unsupervised} applies the prediction uncertainty to filter pseudo-labels during offline PL. The model is first trained on labeled source data for 10k updates. Then, we perform offline PL on target data for 4 iterations and each iteration consists of 5k updates. The target domain LM is used to generate pseudo-labels. To get the uncertainty estimation, we compute pseudo-labels with dropout for 3 times.
\end{itemize}

As shown in the upper part of \autoref{tab:main}, by leveraging additional unlabeled target domain data with the continued pre-training technique, the robust wav2vec model consistently outperforms the wav2vec 2.0 model on all datasets. 

On the basis of the robust wav2vec model, DAT could further improve the performance on CV datasets by encouraging the invariant representation between source and target domain. However, such improvement could not generalize to TED and SWBD datasets. The reason is that the CV dataset and the source domain dataset are both read speech, while TED and SWBD are in different linguistic domains, i.e., lecture and conversational speech. Although DAT is observed to be effective for ASR in previous work, they mostly use it under the acoustic domain shift, e.g., mismatches of environment, device, accent, etc. And the linguistic style of the source and target domain are similar, e.g., both read speech. This phenomenon is in line with the findings in \cite{li2020rethinking}, which indicates DAT is less effective with the label distribution shift. 

On the contrary, both online (MPL) and offline (DUST) PL approaches could effectively boost the cross-domain performance on all datasets, illustrating better generalization ability than DAT. Specifically, MPL clearly outperforms DUST when there is only the acoustic mismatch, i.e., accent mismatch in CV dataset. And DUST has similar or better performance with MPL when there is also the linguistic mismatch, i.e., on TED and SWBD datasets. It illustrates that the online PL is more suitable for tackling the acoustic mismatch while the offline PL efficiently alleviates the linguistic mismatch by utilizing the target domain LM.

By utilizing advanced online and offline PL algorithms, both online and offline PL versions of CASTLE exhibit clear gains over their corresponding previous work. In particular, the online PL version of CASTLE surpasses the performance of robust wav2vec + MPL, and the offline version outperforms robust wav2vec + DUST. Notably, the online PL version already achieves better performance than existing approaches across all datasets, and the two-step version further enhances performance.

We also list some supervised results in the lower part of \autoref{tab:main} to show the topline performance. 
As expected, the supervised results significantly outperform the unsupervised results. The continued pre-training is also shown to be effective in boosting supervised training performance. And using both source and target data for fine-tuning is slightly better than using only target data.  \method effectively closes the performance gap between the supervised and unsupervised approaches. Moreover, \method could achieve comparable performance with supervised approaches on TED dataset, where the domain mismatch is the least.

We further give a direct comparison with results from related literature on the TED dataset. As shown in \autoref{tab:comparision}, our approach significantly outperforms all previously reported unsupervised results and achieved competitive results with some supervised approaches. 

\begin{table}[htbp]
  \centering
  \caption{Direct comparison with previous literature on TED.}
    \begin{tabular}{ccccc}
    \toprule
    \multirow{2}[4]{*}{Method} & \multicolumn{2}{c}{External Data} & \multicolumn{2}{c}{WER\%} \\
\cmidrule(lr){2-3} \cmidrule(lr){4-5} & Labeled & Unlabeled & Dev & Test \\
    \midrule
    \textit{\textbf{Unsupervised results}} &     &     &     &  \\
    DUST \cite{khurana2021unsupervised} & 80h & ×   &     & 17.6 \\
    MPL \cite{higuchi2021momentum} & 100h & ×   & 16.2 & 14.9 \\
    Robust wav2vec\cite{hsu2021robust} & 10h & 950h & 8.9 & - \\
    CASTLE & 100h & 860h & \textbf{6.6} & \textbf{6.3} \\
    \midrule
    \midrule
    \textit{\textbf{Supervised results}} &     &     &     &  \\
    ESPnet-Conformer \cite{guo2021recent} & ×   & ×   & 9.6 & 7.6 \\
    UniSpeech \cite{wang2021unispeech} & 960h & ×   & 7.6 & 7.6 \\
    SpeechStew \cite{chan2021speechstew} & 4700h & ×   & -   & 5.3 \\
    SOTA \cite{likhomanenko2020rethinking} & 9000h & ×   & \textbf{5.0} & \textbf{4.7} \\
    \bottomrule
    \end{tabular}%
  \label{tab:comparision}%
\end{table}%

In the following sections, we perform detailed experiments to analyze each component in \method, where the most important ones contributing to the performance are: DPL, UCF and two-step PL.

\subsection{Results of Continued Pre-training with Data Replay}
In this section, we examine different continued pre-training strategies. Specifically, we compare (1) the source domain wav2vec 2.0 model, (2) continued pre-training the source model on target data, and (3) continued pre-training on target data while replaying source data. The continued pre-training updates for (2) and (3) are 20k. And the data replay ratio $p_{s/t}$ between source and target data is 0.5 for (3).

We evaluate pre-trained models with three fine-tuning strategies: (1) source-only: fine-tuning only on the labeled source data for 30k updates; (2) online PL: fine-tuning on both domains with the DPL approach for 20k updates; (3) offline PL: first fine-tuning on the labeled source data for 10k updates and then fine-tuning on the unlabeled target data with UCF-based offline PL for 4 iterations, where each iteration consists of 5k updates and target domain LM is always used when generating pseudo-labels.

\begin{table}[htbp]
  \centering
  \caption{Effectiveness of continued pre-training strategies}
    \begin{tabular}{lccc}
    \toprule
    \multirow{2}[4]{*}{Pre-Train Method} & \multicolumn{3}{c}{WER\% on RT03, Fine-Tune with} \\
\cmidrule{2-4}          & Source-Only & Offline PL & Online PL \\
    \midrule
    Wav2vec 2.0 & 35.9  &  21.8  & 19.7 \\
    + Continued pre-train & 24.8  & 17.9  & \textbf{16.9} \\
    ++ Data replay & \textbf{22.0} & \textbf{17.3} & 17.7 \\
    \bottomrule
    \end{tabular}%
  \label{tab:CPT-DR}%
\end{table}%

As shown in \autoref{tab:CPT-DR}, continued pre-training consistently improves the performance for three fine-tuning strategies. The data replay technique further enhances the performance of the source-only fine-tuning by avoiding the catastrophic forgetting of source knowledge. And the offline PL performance has the same trend since source-only fine-tuning determines the quality of the seed model in offline PL. On the contrary, the online PL works better without data replay. We hypothesize that the reason is that simultaneously training on both domains could alleviate the catastrophic forgetting issue. Therefore, given a certain training update, continued pre-training on only the target domain leads to better target domain ability and better PL training on the target domain.

\begin{figure}[htbp]
    \centering
	\subfloat[Selected Proportion] 
	{ \label{fig:two-stage_confidence_selected}
		\includegraphics[width=0.65\columnwidth]{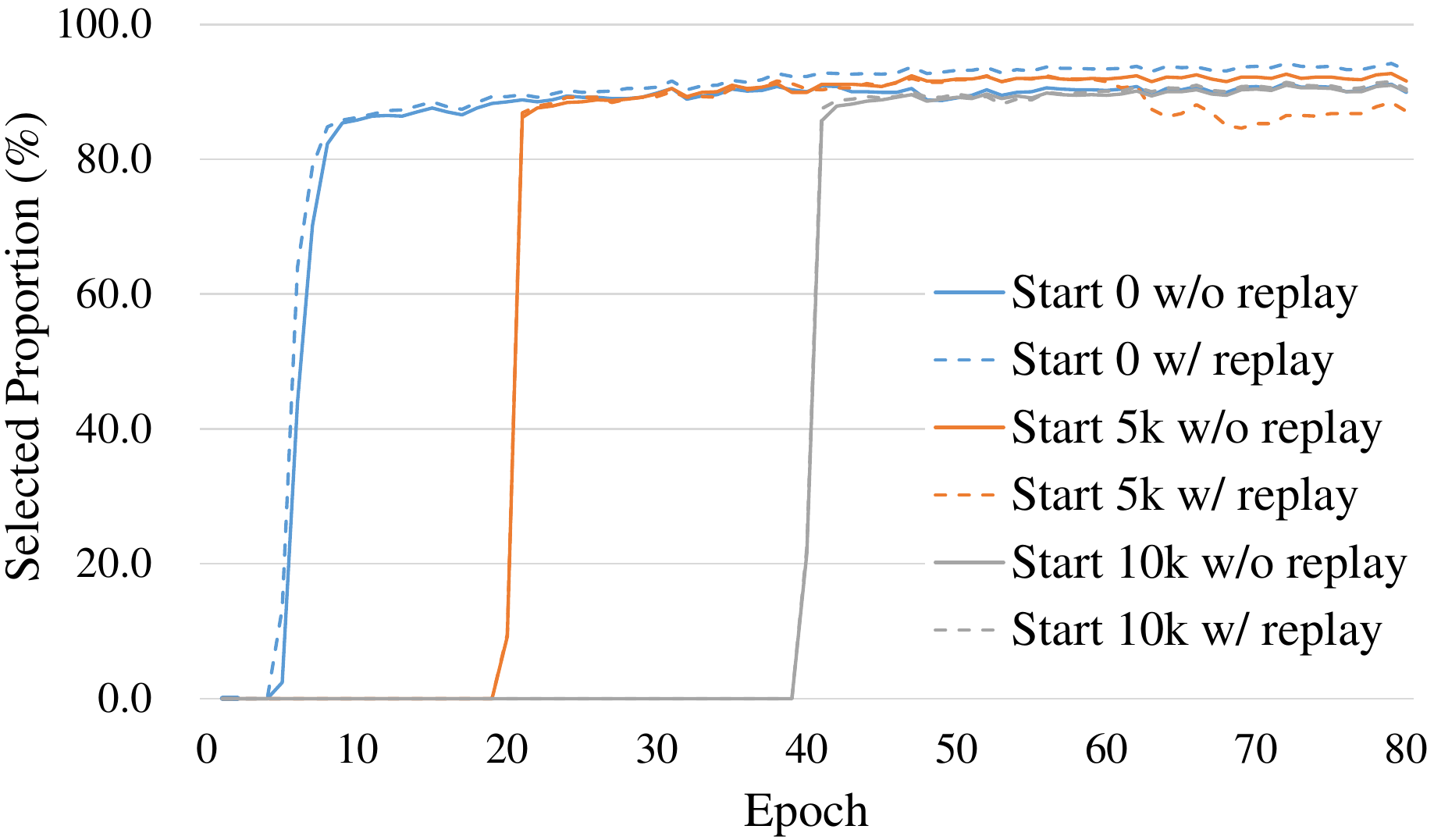}
	}
	
	\subfloat[Evaluation WER] 
	{ \label{fig:two-stage_confidence_wer}
		\includegraphics[width=0.65\columnwidth]{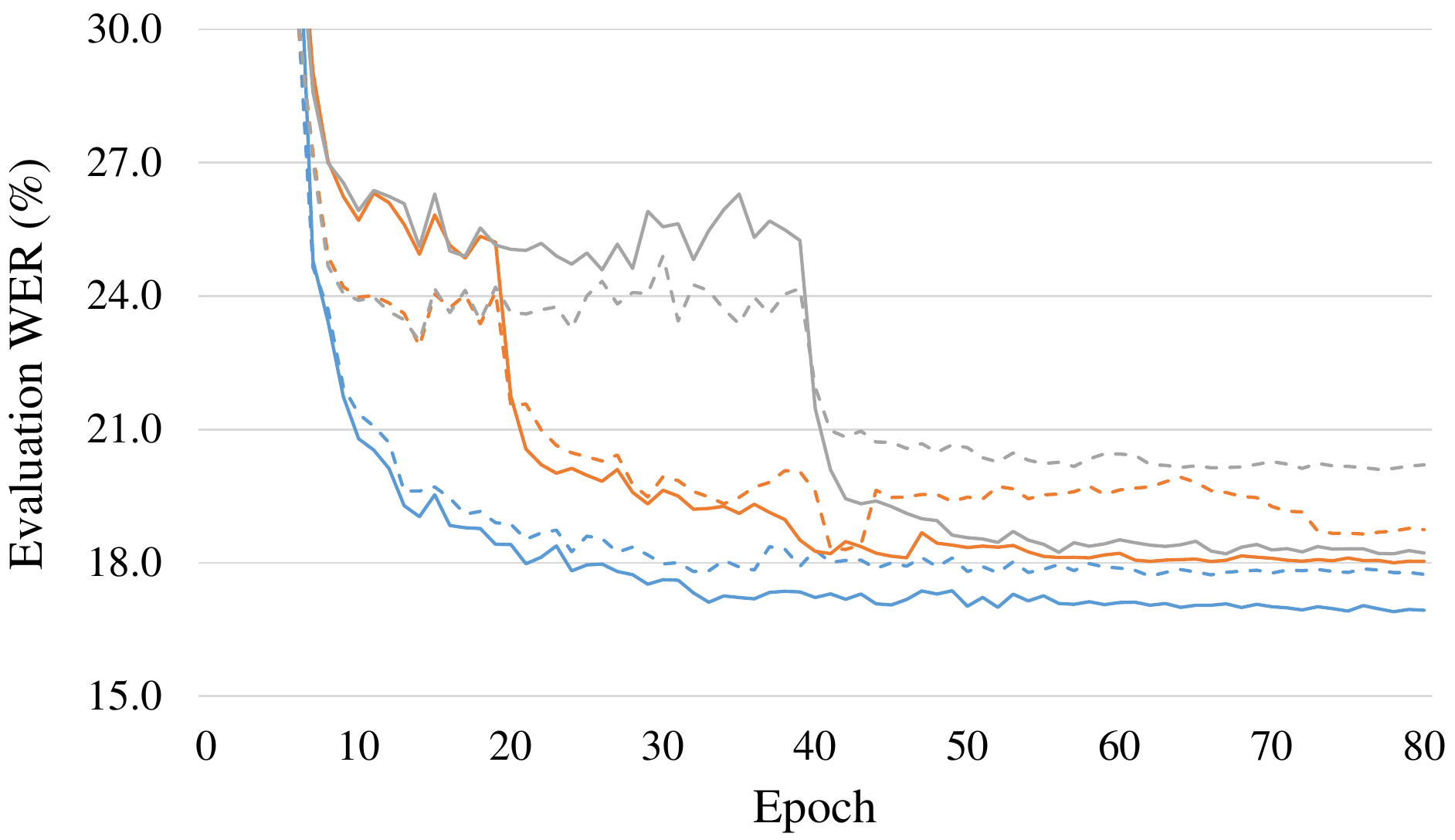}
	}
	\caption{Two-stage online PL behavior over epochs with different continued pre-training strategies and start updates. (a) Selected proportions of the unlabeled target domain training set. (b) Evaluation WER on the RT03 development set.} 
	\label{fig:two-stage_DPL}
\end{figure}

We verify the hypothesis with the two-stage DPL training, where the first stage is trained only with labeled source data and the PL loss is added in the second stage. As shown in \autoref{fig:two-stage_DPL}, we conduct three sets of comparisons with different start update numbers (0, 5k, 10k) of the second stage. The model with data replay is consistently better in the first stage. In the second stage, the one without data replay quickly catches up and achieves better results. This phenomenon verifies that the model without data replay works better as long as the model is simultaneously fine-tuned on both domains.

In conclusion, different fine-tuning strategies require different continued pre-training strategies. When fine-tuning with source-only data, the SSL pre-trained model must incorporate knowledge of both domains to generalize to the target domain. Thus data replay could benefit source-only fine-tuning and offline PL. On the other hand, online PL does not suffer from the catastrophic forgetting of source knowledge and could benefit from concentrating on the target domain, i.e., without data replay. Therefore, \method does not utilize data replay since the online PL approach DPL is used to fine-tune the SSL pre-trained model.

\subsection{Results of Online PL with DPL}
\label{sec:online_PL}

In this section, we conduct experiments to show the effectiveness of DPL. Firstly, we compare DPL with two other online PL approaches: vanilla online PL \cite{chen2020semi} and MPL \cite{higuchi2021momentum}.  All three approaches start from a continued pre-trained model without data replay and the fine-tuning updates are 20k. The hyper-parameters are tuned on the development set. Note that the two-stage training does not clearly improve the vanilla online PL. Thus we apply the single-stage training for it.

\begin{table}[htbp]
  \centering
  \caption{Comparison of different online PL approaches}
    \begin{tabular}{lccc}
    \toprule
    \multirow{2}[4]{*}{PL Method} & \multicolumn{3}{c}{WER\%} \\
\cmidrule{2-4}        & RT03 & H-SB & H-CH \\
    \midrule
    Vanilla online PL & 19.1 & 12.5 & 19.6 \\
    MPL   & 18.6  & 13.5  & 18.3 \\
    DPL & \textbf{16.9} & \textbf{11.7} & \textbf{17.6} \\
    - Dual-branch & 17.7 & 12.2 & 18.6 \\
    - Confidence filter & 17.6 & 12.1 & 18.3 \\

    \bottomrule
    \end{tabular}%
  \label{tab:online_PL_comparison}%
\end{table}%

\begin{figure*}[htbp]
    \centering
	\subfloat[Selected Proportion] 
	{ \label{fig:confidence_selected}
		\includegraphics[width=0.55\columnwidth]{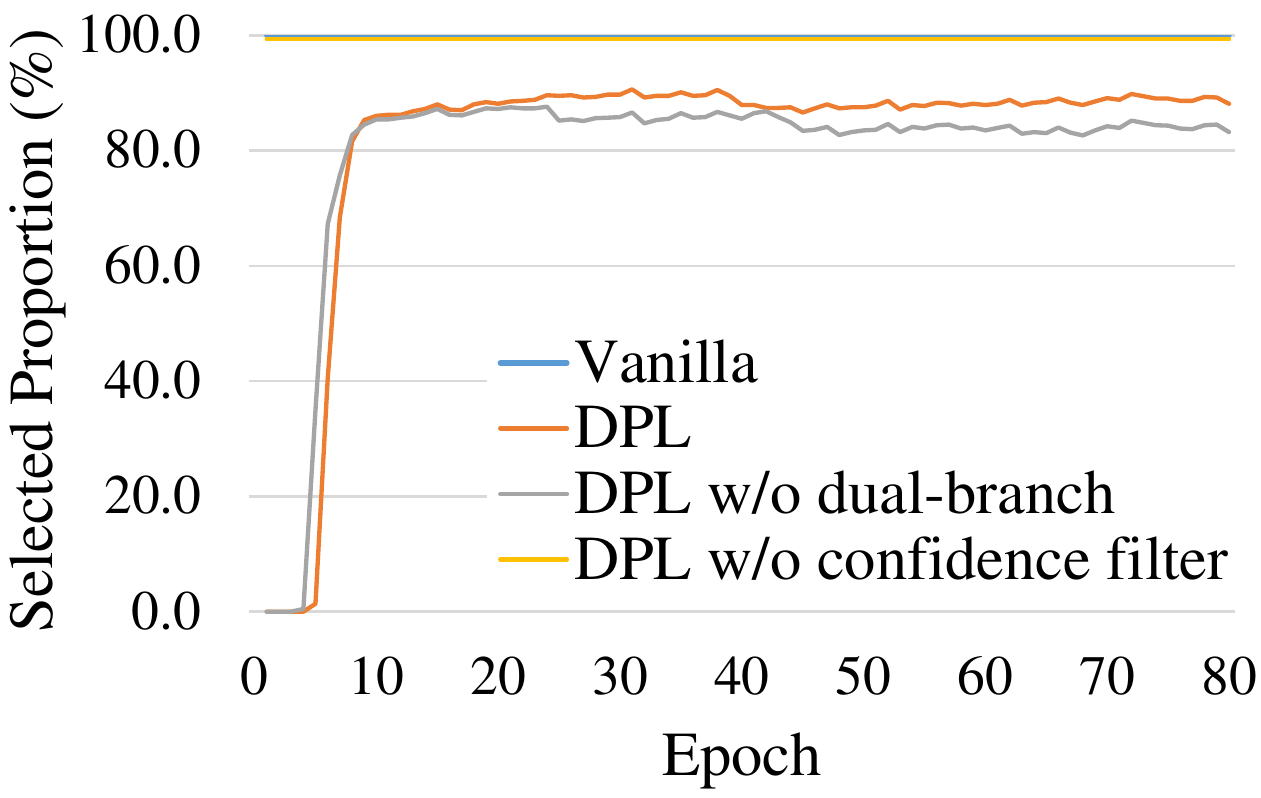}
	}\quad
	\subfloat[Selected WER] 
	{ \label{fig:confidence_selected_wer}
		\includegraphics[width=0.55\columnwidth]{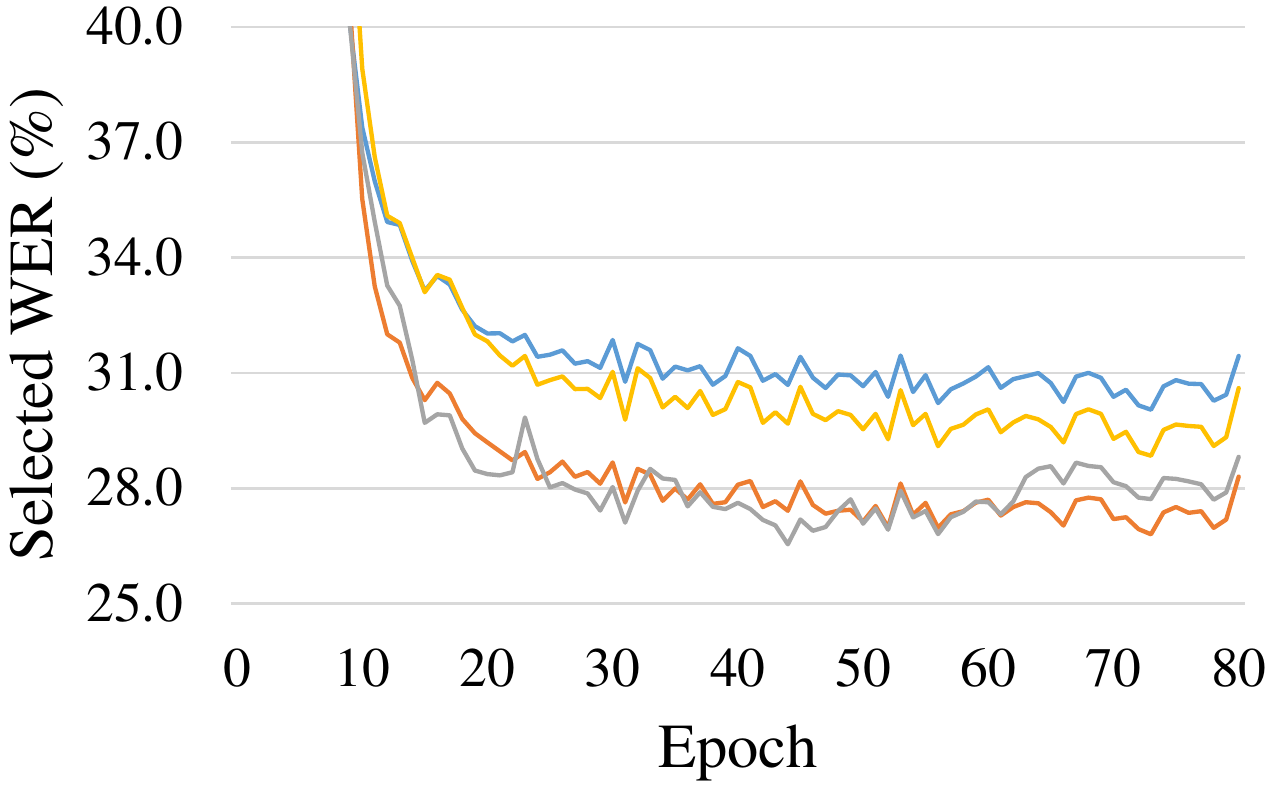}
	}\quad
	\subfloat[Evaluation WER] 
	{ \label{fig:confidence_wer}
		\includegraphics[width=0.55\columnwidth]{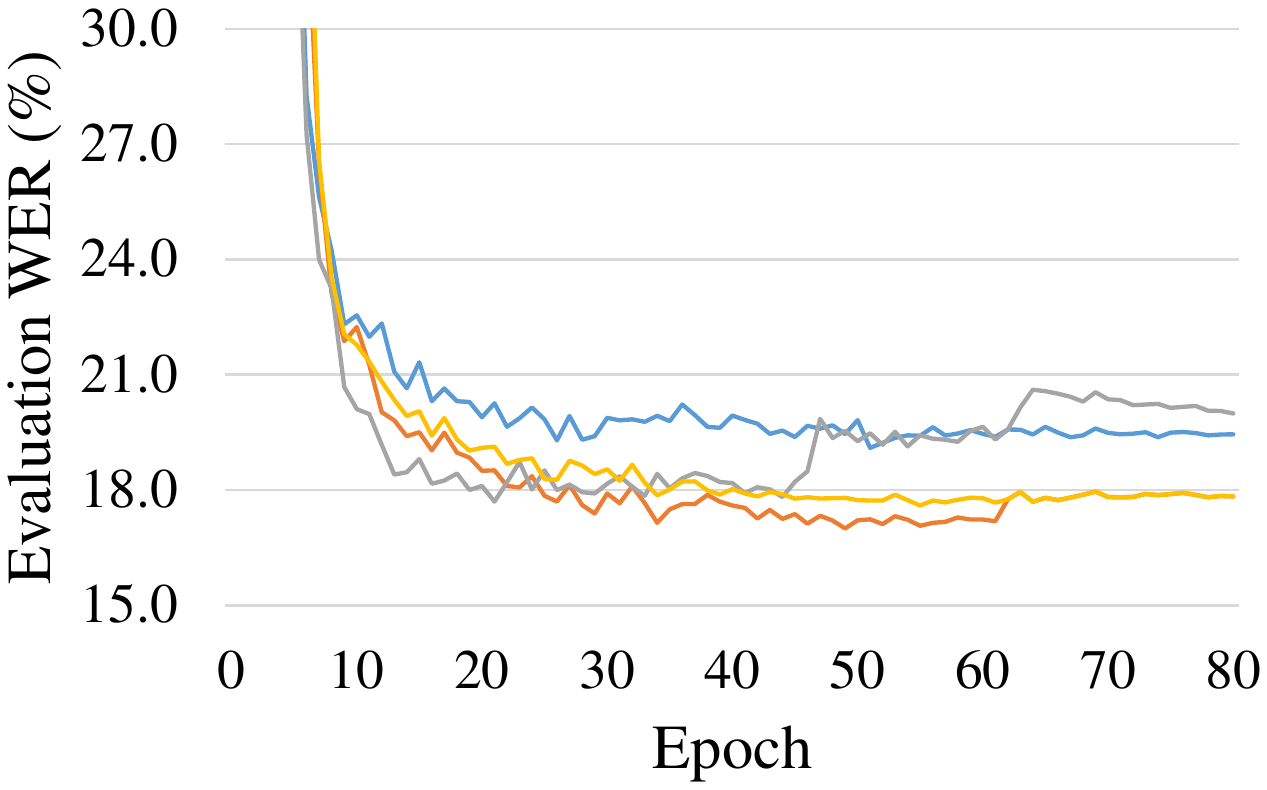}
	}
	\caption{Behaviors of different online PL approaches over epochs. (a) Selected proportions of the unlabeled target domain training set. (b) The WER of the selected training set. (c) Evaluation WER on the RT03 development set.} 
	\label{fig:confidence}
\end{figure*}

As shown in \autoref{tab:online_PL_comparison}, DPL significantly outperforms other two approaches. Compared with the vanilla online PL, DPL involves two essential components: dual-branch learning and confidence filtering. We conduct the ablation study for DPL and find the performance is worse if any component is removed. Furthermore, we illustrated the selected proportion and WER of pseudo-labels, as well as the evaluation WER on the development set in \autoref{fig:confidence}. Comparing DPL and DPL w/o confidence filter, the confidence filtering technique effectively reduces WER by filtering out the erroneous pseudo-labels. Comparing DPL and DPL w/o dual-branch, in the earlier updates, the one w/o dual-branch has a similar selected proportion and slightly better selected pseudo-labels, thus giving a better performance. Then, in the later updates, the one w/o dual-branch suffers from the error accumulation issue, leading to obvious performance degradation. In the meantime, the one w/ dual-branch gradually catches up and leads to better performance.

\begin{table}[htbp]
  \centering
  \caption{Variants of DPL}
    \begin{tabular}{lccc}
    \toprule
    \multirow{2}[4]{*}{PL Method} & \multicolumn{3}{c}{WER\%} \\
\cmidrule{2-4}          & RT03  & H-SB  & H-CH \\
    \midrule
    DPL   & \textbf{16.9} & \textbf{11.7} & \textbf{17.6} \\
    + Pseudo-label from main branch & 17.9  & 12.0  & 18.5 \\
    + Two-stage Training & 16.8  & 11.5  & 17.8 \\
    + EMA (discount factor = 0.001) & 19.4  & 13.7  & 19.8 \\
    + Two-layer auxiliary branch & 17.6  & 12.1  & 18.0 \\
    \bottomrule
    \end{tabular}%
  \label{tab:variant_DPL}%
\end{table}%

To further understand DPL, we show the results of some variants of DPL in \autoref{tab:variant_DPL}. Firstly, we train a model that uses the auxiliary branch as a regularization. And the pseudo-labels are still from the main branch, like the vanilla online PL. The performance degradation of this model shows that generating pseudo-labels with the auxiliary branch is the key to the success of DPL. Secondly, instead of using confidence filtering to determine the utilized pseudo-labels automatically, we try the two-stage training where we manually choose when we start to use the PL loss. Like the vanilla online PL, DPL also does not benefit from the two-stage training. Thirdly, we add the EMA technique to DPL. We find a larger discount factor~\cite{manohar2021kaizen} (0.01 or 0.1), which gives more weight to the recent parameters, does not make a significant difference. And a smaller discount factor (0.001) slows down the convergence speed and leads to worse performance. Therefore, EMA is not necessary when the PL training is already stable. It is in line with the findings in \cite{manohar2021kaizen}, which illustrates that EMA could stabilize the PL training but lead to slow improvement. Lastly, we explore using more layers in the auxiliary branch. Specifically, we utilized one more transformer layer in addition to the last linear layer. The result shows that using only one linear layer is enough and performs better in practice.

\begin{figure}[htbp]
	\centering
	\includegraphics[width=0.55\columnwidth]{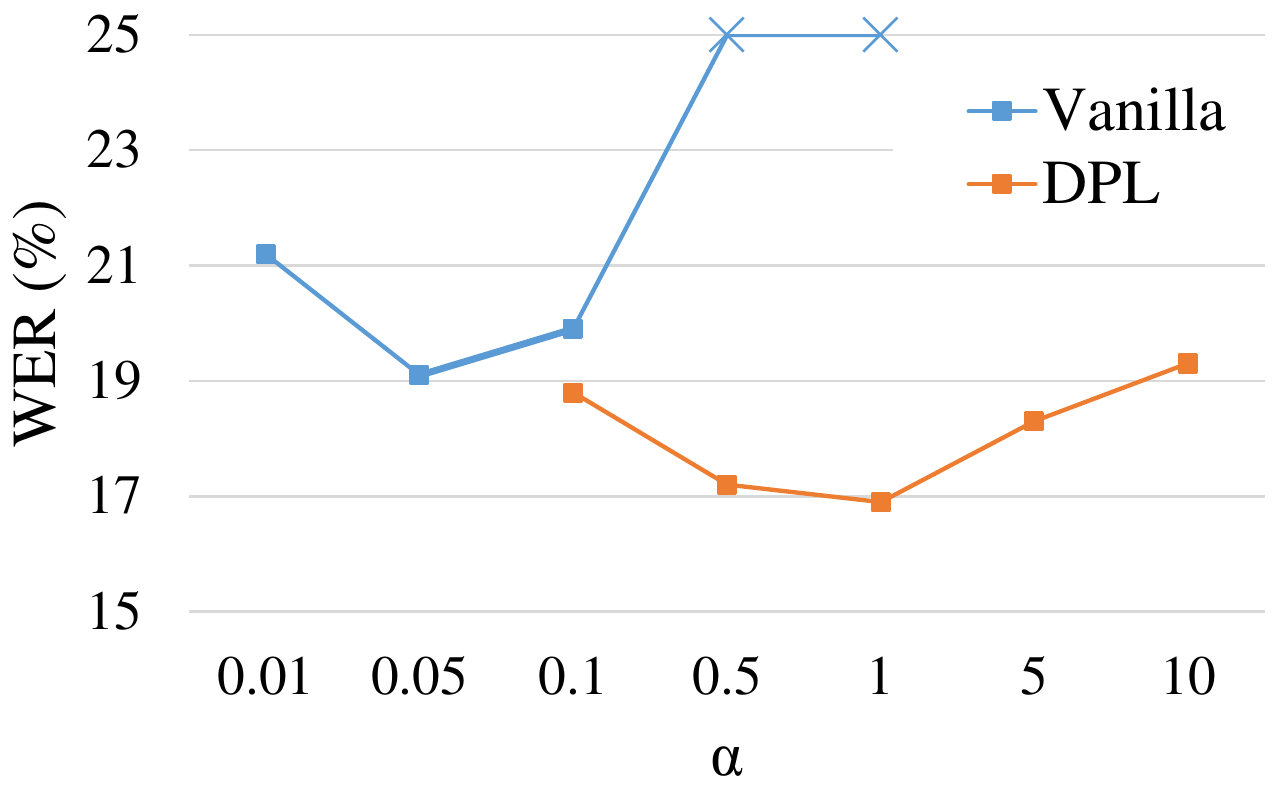}
	\caption{Performance of different online PL approaches and different PL loss weights ($\alpha$) on the RT03 development set. The cross symbol denotes the collapsed model trained to the trivial solution.} 
	\label{fig:different_alpha}
\end{figure}

Next, we evaluate the DPL's sensitivity to the PL loss weight $\alpha$ in \autoref{equ:DPL}. For comparison, we also evaluate the vanilla online PL. As shown in \autoref{fig:different_alpha}, when $\alpha$ is large, the vanilla online PL could lead to the trivial solution where the model predicts blank for any input. On the other hand, DPL is more stable and leads to consistently better performance.

Given the collapsed vanilla online PL training setup ($\alpha=0.5$) in \autoref{fig:different_alpha}, we illustrate how the confidence score in \autoref{equ:confidence} could avoid the trivial solution. Since the proposed confidence score specifically discards the frames where the maximum score belongs to the blank, we also compare it with the variant that keeps blank in the computation.  

\begin{figure*}[htbp]
    \centering
	\subfloat[Selected Proportion] 
	{ \label{fig:trival_selected_proportion}
		\includegraphics[width=0.65\columnwidth]{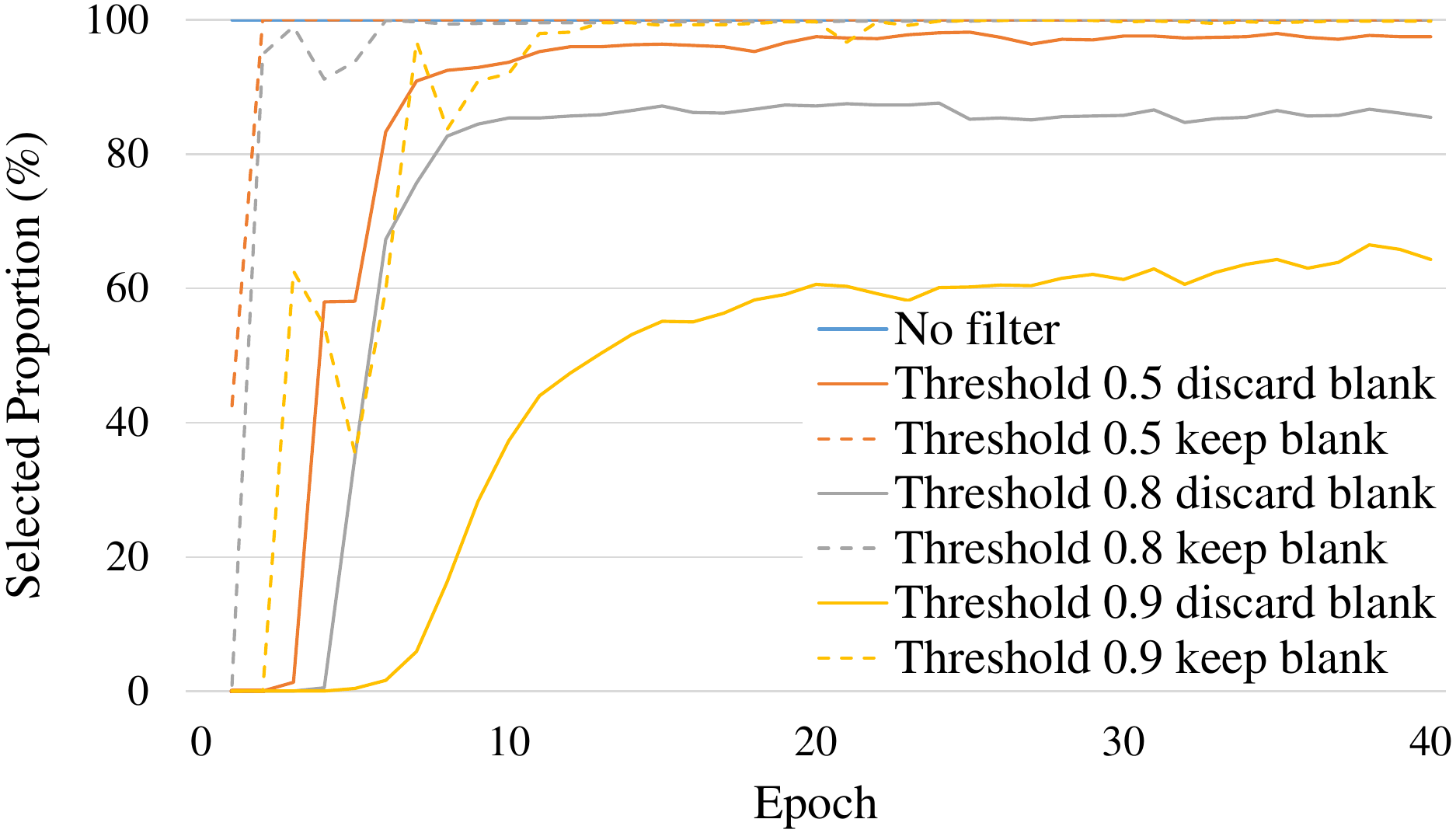}
	}
	\subfloat[Selected WER] 
	{ \label{fig:trival_selected_wer}
		\includegraphics[width=0.65\columnwidth]{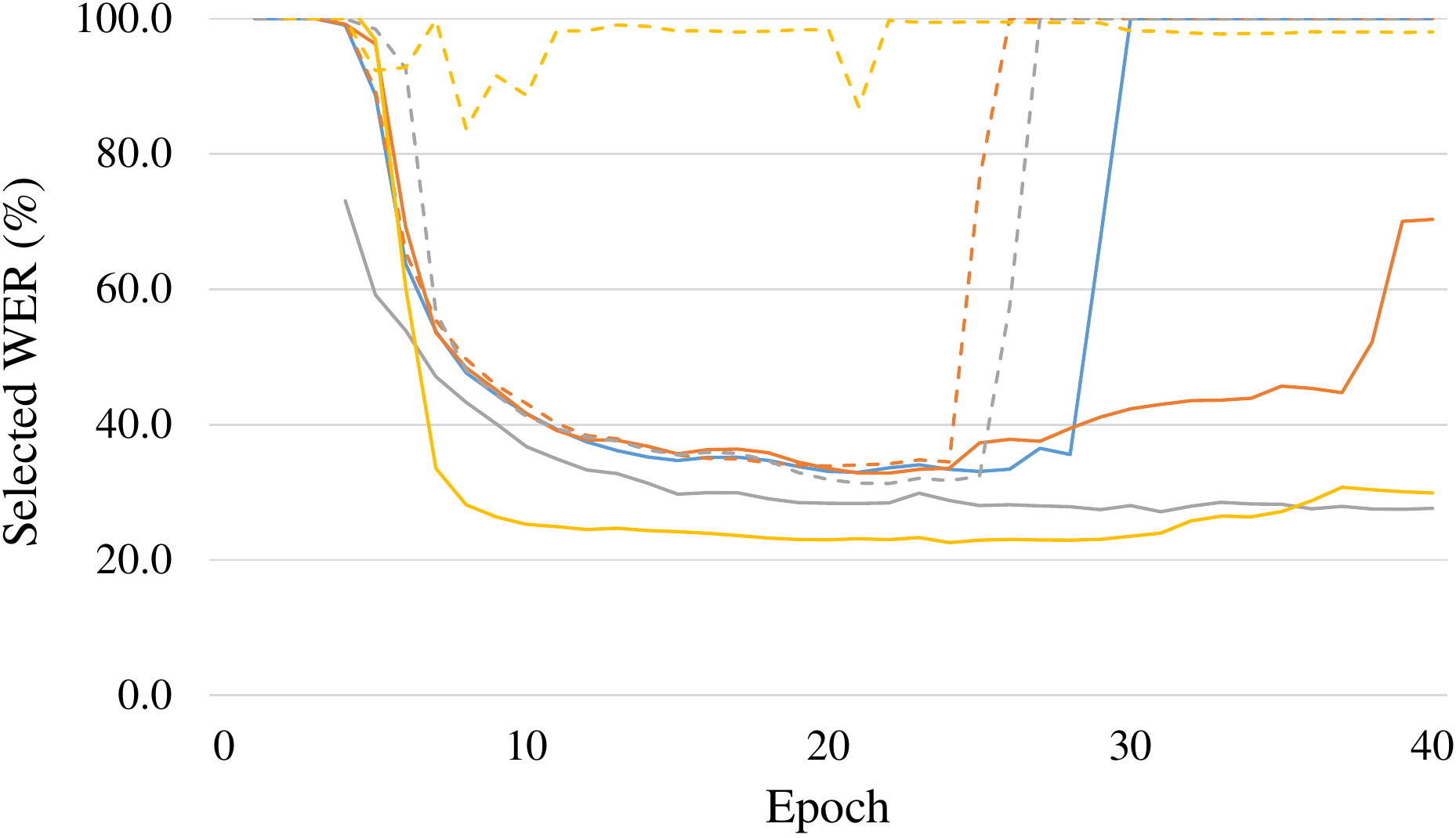}
	}
	\subfloat[Evaluation WER] 
	{ \label{fig:trival_evaluation_wer}
		\includegraphics[width=0.65\columnwidth]{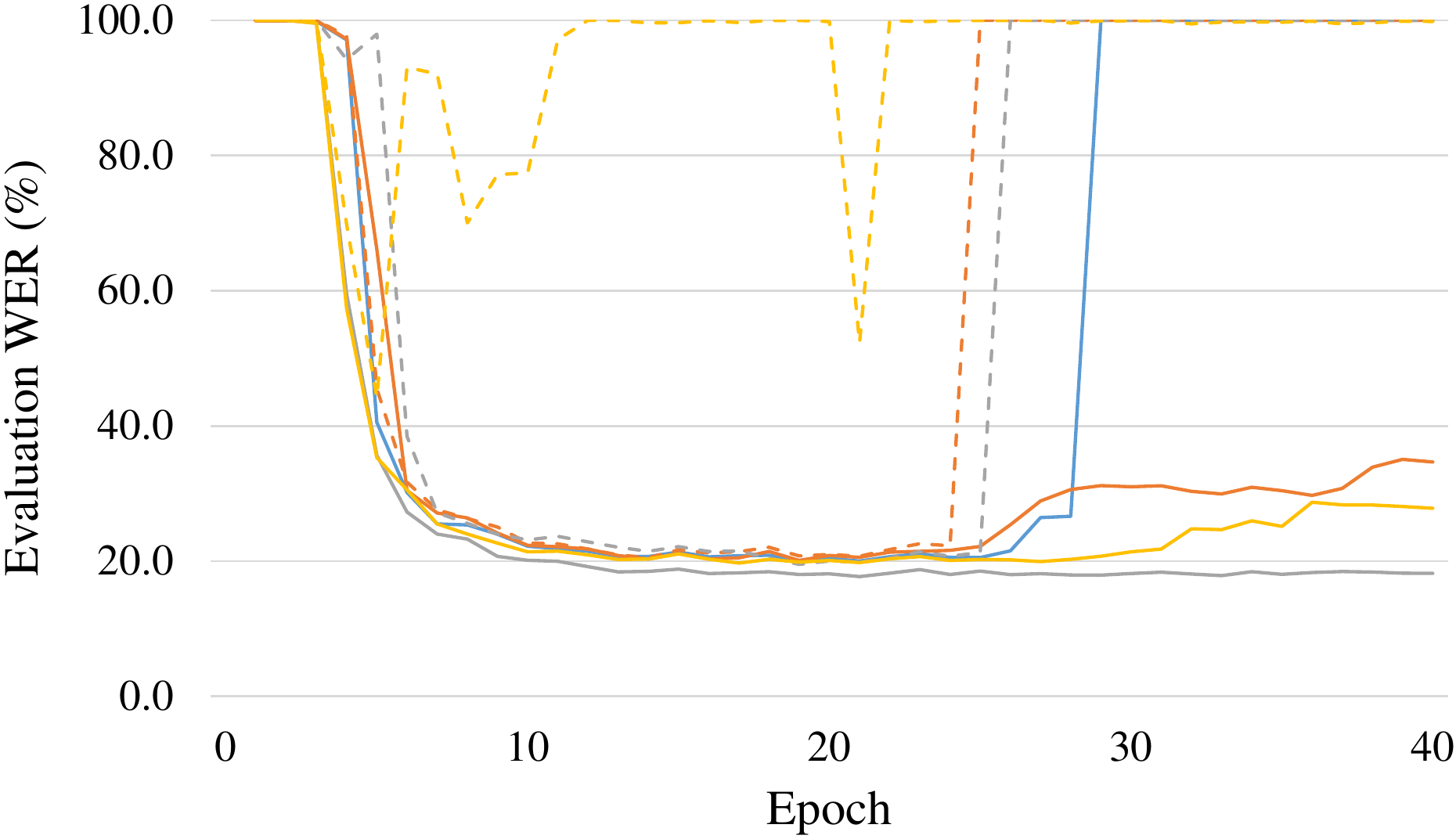}
	}
	\caption{Behaviors of vanilla online PL over epochs with different filtering strategies and confidence thresholds. (a) Selected proportions of the unlabeled target domain training set. (b) The WER of the selected training set. (c) Evaluation WER on the RT03 development set.} 
	\label{fig:trival}
\end{figure*}

As shown in \autoref{fig:trival}, when no filtering strategy is applied, all pseudo-labels are used in training. The blank (or mostly blank) pseudo-labels can mislead the model to the trivial solution. Therefore, the selected WER and the evaluation WER quickly increase to 100\% at a point.

Then, we examine how confidence filtering affects the result. If we keep blanks when computing the confidence score, the blank (or mostly blank) pseudo-labels will have a high confidence score and will be kept in the selected subset. Consequently, the blank pseudo-labels would be more dominating in the selected subset, and the trivial solution happens even earlier than the one with no filtering.
On the contrary, the confidence score that discards blank could benefit from more accurate selected pseudo-labels without the trouble of the blank pseudo-labels, thus being more stable and leading to better performance.

 \subsection{Results of Offline PL with Different Filtering Strategies}
 \label{sec:offline_PL}
In this section, we conduct experiments to compare different filtering strategies for offline PL. Specifically, we consider confidence filtering~\cite{kahn2020self}, uncertainty filtering~\cite{khurana2021unsupervised}, and the proposed UCF. We first generate pseudo-labels with beam-search decoding and target domain LM. Then, we select some fixed proportions (10\%, 30\%, 50\%, 70\%, 100\%) of the pseudo-labels with these filtering approaches and compute the WER of the selected subsets. Finally, we fine-tune the unadapted or the adapted model on the selected subsets for one iteration (5k updates) and evaluate the WER on the RT03 development set. Note that the unadapted model is the model fine-tuned only on source data on the basis of the source domain pre-trained model. And the adapted model is the model fine-tuned on both source and target data with online PL on the basis of the continued pre-trained model.

\begin{figure*}[!t]
    \centering
	\subfloat[Selected WER (unadapted)] 
	{ \label{fig:unadapted_selected_wer}
		\includegraphics[width=0.5\columnwidth]{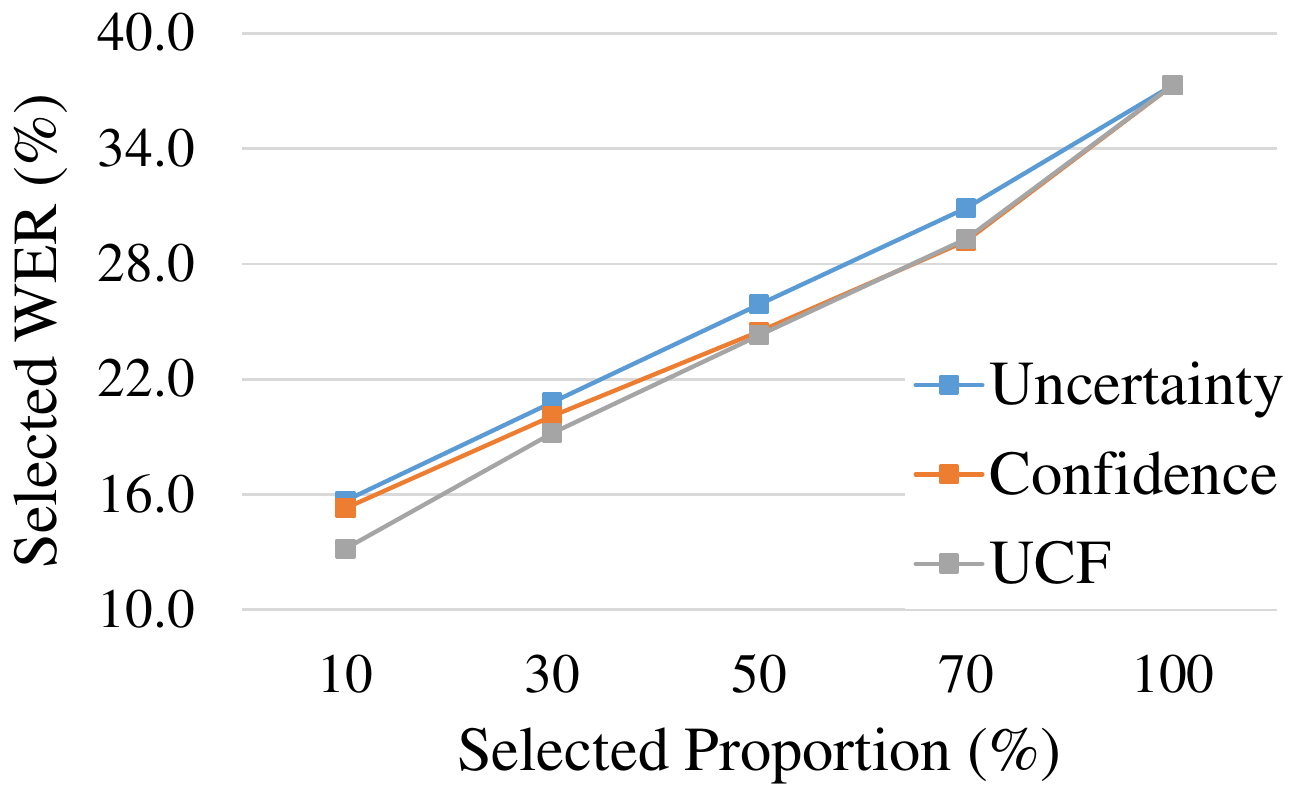}
	}
	\subfloat[Evaluation WER (unadapted)] 
	{ \label{fig:unadapted_evaluation_wer}
		\includegraphics[width=0.5\columnwidth]{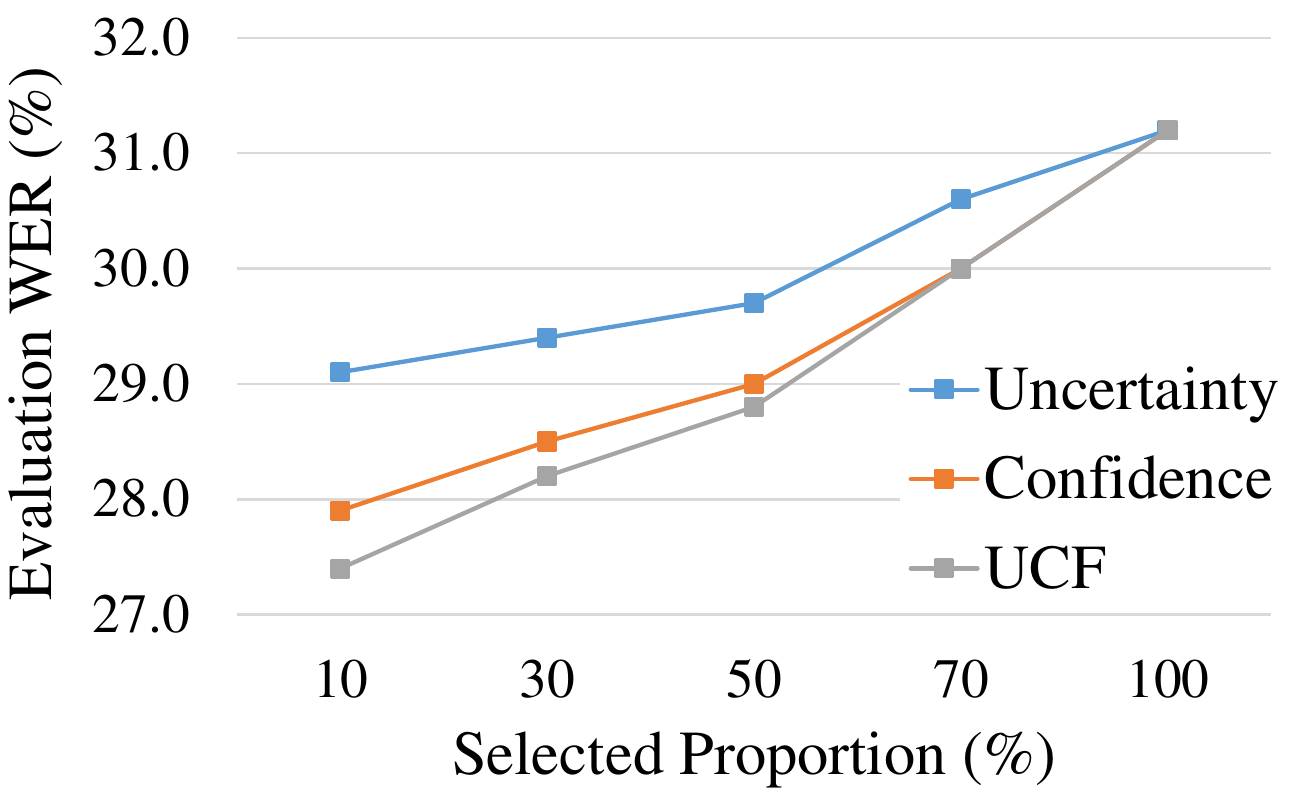}
	}
	\subfloat[Selected WER (adapted)] 
	{ \label{fig:adapted_selected_wer}
		\includegraphics[width=0.5\columnwidth]{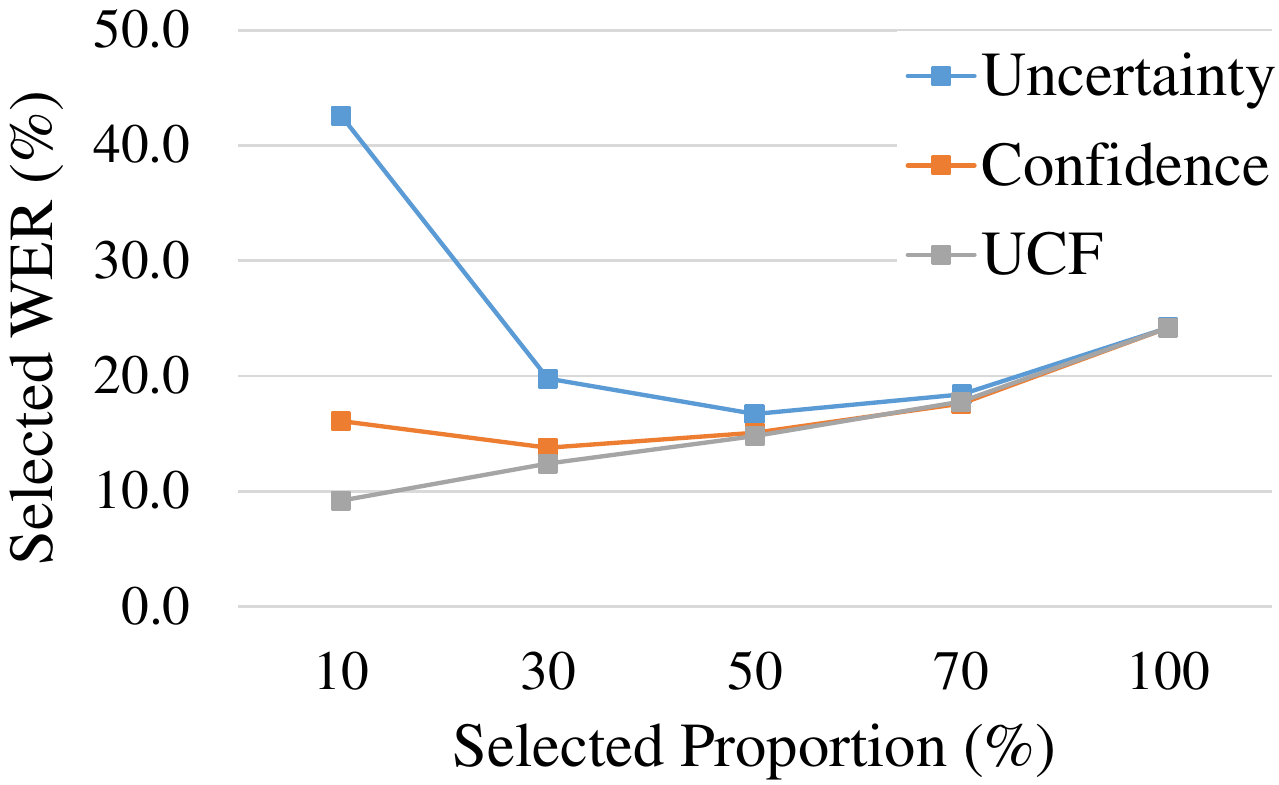}
	}
	\subfloat[Evaluation WER (adapted)] 
	{ \label{fig:adapted_evaluation_wer}
		\includegraphics[width=0.5\columnwidth]{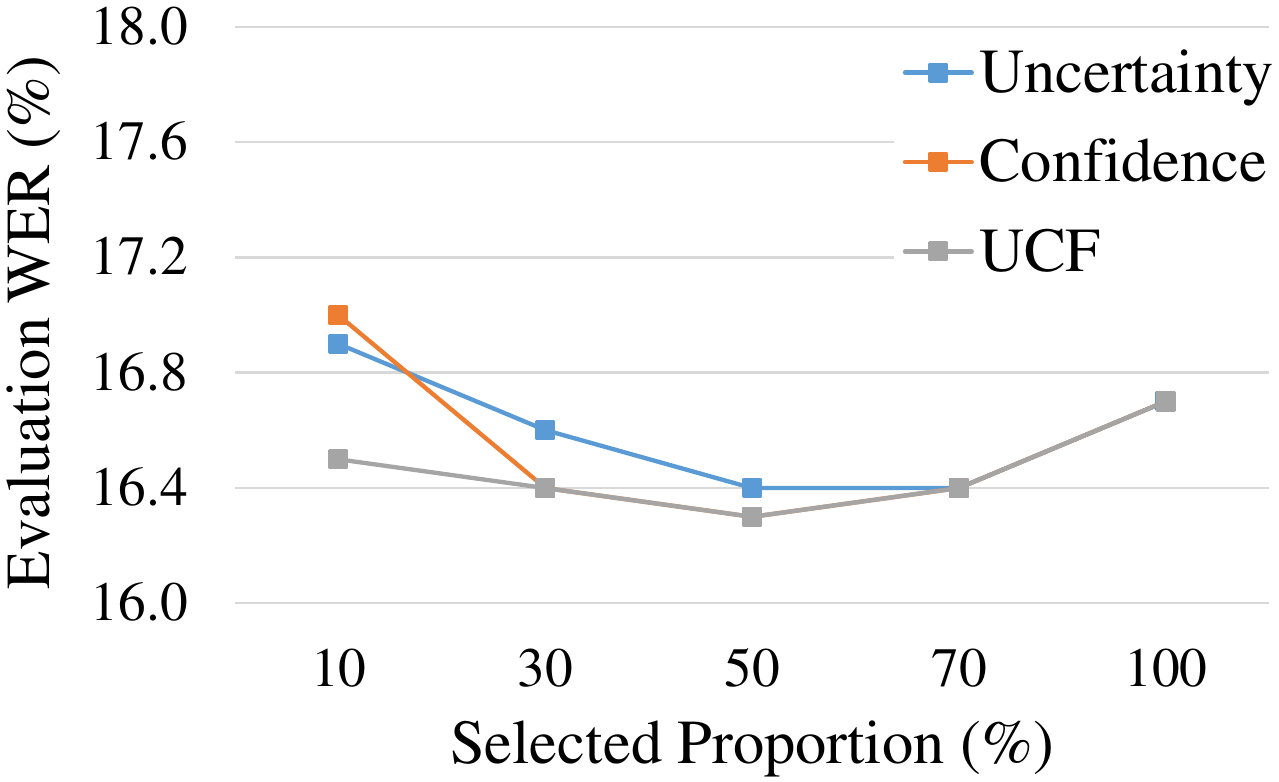}
	}
	\caption{Behaviors of the unadapted and adapted models with different filtering strategies for offline PL. (a) and (c) are the WER of the selected training set. (b) and (d) are evaluation WER on the RT03 development set.} 
	\label{fig:unadapted_filter}
\end{figure*}

As shown in \autoref{fig:unadapted_filter}, for the unadapted model, since there are lots of errors in the pseudo-labels, no matter which filtering approach is used, the fewer pseudo-labels we select, the lower WER is obtained on the selected training subset. And the evaluation WER of the fine-tuned model is also decreased correspondingly. Among these filtering strategies, UCF is consistently better than others. 

As for the adapted model (shown in \autoref{fig:unadapted_filter}), the confidence filtering and the uncertainty filtering could lead to a higher WER when we select a small proportion (10\%) of pseudo-labels. And the evaluation WER increases correspondingly. This phenomenon indicates that the most certain or the most confident predictions are not the most accurate ones. On the contrary, UCF could effectively alleviate this issue by leveraging both criteria and estimating the hyper-parameters on the development set. Consequently, UCF significantly outperforms the other two filtering approaches when selecting a small proportion of pseudo-labels.
The best evaluation WER is achieved when we select 50\% of pseudo-labels with UCF or confidence filtering since the pseudo-labels from the adapted model are more accurate than the unadapted model. 

\subsection{Results of Two-Step PL}
\label{sec:two_step_PL}

In this section, we evaluate the effectiveness of the two-step PL. For comparison, we list the performance of the one-step PL approaches, i.e., online or offline PL. Note that all approaches are trained with a total of 30k updates. To present the best results of the three approaches, the offline PL utilizes the data replay technique during continued pre-training, while the other two approaches do not.

\begin{table}[htbp]
  \centering
  \caption{Comparison between two-step PL and one-step online/offline PL. Source or target domain LM is used to generate pseudo-labels.}
    \begin{tabular}{lcccc}
    \toprule
    \multirow{2}[4]{*}{PL Method} & \multirow{2}[4]{*}{LM Domain} & \multicolumn{3}{c}{WER\%} \\
\cmidrule{3-5}        &     & RT03 & H-SB & H-CH \\

    \midrule
    Online PL & -   & 17.0 & 11.8 & 17.6 \\
    \midrule
    Offline PL & Target & 17.3 & 11.2 & 17.6 \\
        \midrule
    \multirow{2}[1]{*}{Two-step PL} & Source & 16.4 & 11.2 & 17.4 \\
        & Target & \textbf{16.1} & \textbf{10.8} & \textbf{17.1} \\
    \bottomrule
    \end{tabular}%
  \label{tab:two-step_PL}%
\end{table}%

As shown in \autoref{tab:two-step_PL}, the two-step PL consistently outperforms both online and offline PL. Comparing two-step PL and online PL, the two-step PL improves the performance by generating more accurate pseudo-labels for refinement. Both source and target domain LM are beneficial and the target domain LM is better since it could transfer the target domain linguistic knowledge into the model. Note that the online PL does not benefit from more training update as the 30k updates gives similar results with the 20k updates in \autoref{tab:online_PL_comparison}.

\begin{table}[htbp]
  \centering
  \caption{Generate pseudo-labels with decoding method in online PL w/o confidence filtering.}
    \begin{tabular}{lcccc}
    \toprule
    \multirow{2}[4]{*}{Decode Method} & \multirow{2}[4]{*}{LM weight} & \multicolumn{3}{c}{WER\%} \\
\cmidrule{3-5}        &     & RT03 & H-SB & H-CH \\
    \midrule
    Greedy-decode & -   & 17.6 & 12.1 & 18.3 \\
    \midrule
    \multirow{3}[1]{*}{Beam-search} & 0   & 17.8 & 12.0  & 18.5 \\
        & 0.5 & 17.9 & 12.1 & 18.9 \\
        & 1.0 & 20.1 & 13.2 & 20.8 \\
    \bottomrule
    \end{tabular}%
  \label{tab:online_decode}%
\end{table}%

\begin{table}[htbp]
  \centering
  \caption{Generate pseudo-labels with decoding method in offline PL.}
    \begin{tabular}{lcccc}
    \toprule
    \multirow{2}[4]{*}{Decode Method} & \multirow{2}[4]{*}{LM weight} & \multicolumn{3}{c}{WER\%} \\
\cmidrule{3-5}        &     & RT03 & H-SB & H-CH \\
    \midrule
    \multirow{3}[2]{*}{Beam-search} & 0   & 19.8 & 13.8 & 19.8 \\
        & 0.5 & 17.5 & 11.8 & 17.9 \\
        & 1.0 & 17.3 & 11.2 & 17.6 \\
    \bottomrule
    \end{tabular}%
  \label{tab:offline_decode}%
\end{table}%

Aside from two-step PL, there are two other ways to take advantage of LM: (1) using LM-based decoding to generate pseudo-labels during online PL; (2) utilizing offline PL from the seed model that is only supervised fine-tuned on source labeled data, i.e., pure offline PL w/o online PL. 

In terms of the online PL, as shown in \autoref{tab:online_decode}, we empirically find that utilizing LM-based beam-search decoding to generate pseudo-labels in online PL can not lead to proper improvement. Previous studies~\cite{higuchi2021momentum,likhomanenko2020slimipl} have also opted for LM-free decoding in online PL with the concern that LM-based decoding might lead to over-fitting to LM. Therefore, we seek help from the offline PL to use LM, i.e., two-step PL. When LM is not used, greedy-decoding provides comparable performance with beam-search but has less computation complexity, thus being a suitable choice for pseudo-label generation in online PL.

\begin{figure}[htbp]
	\centering
	\includegraphics[width=0.55\columnwidth]{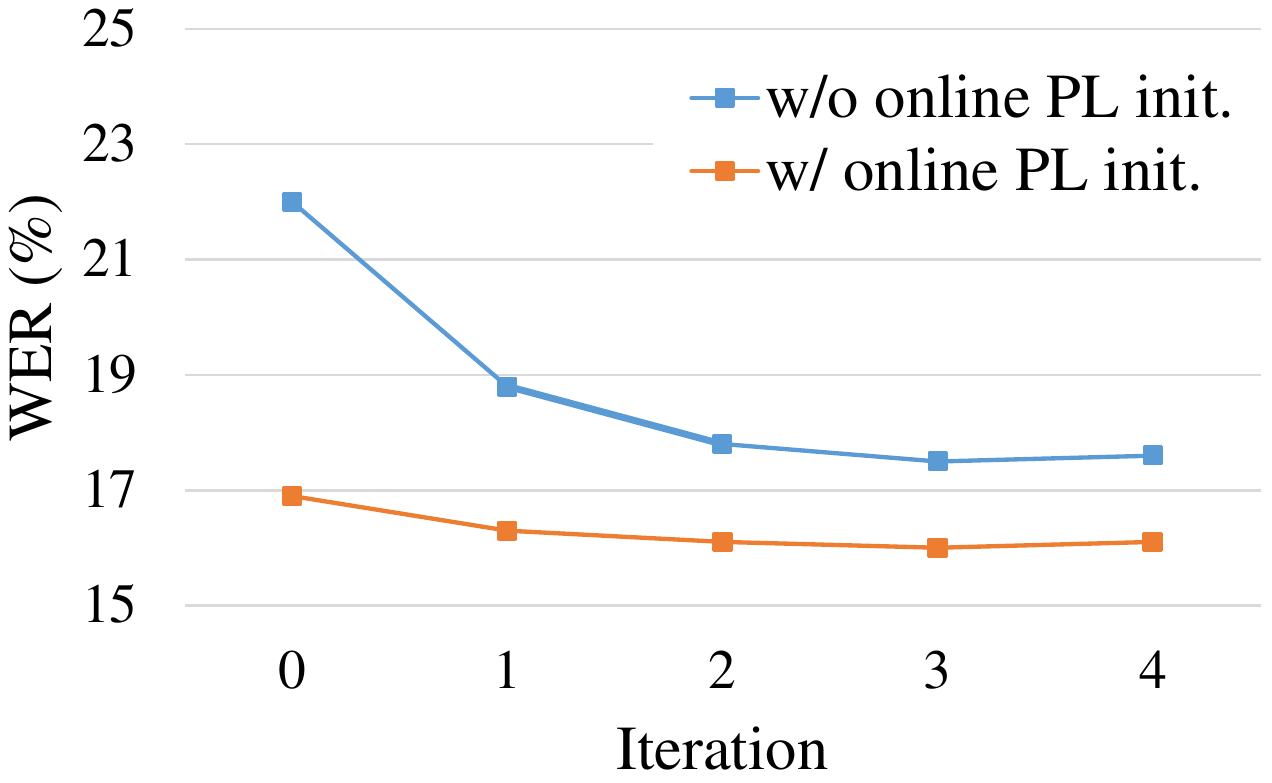}
	\caption{Performance of offline PL over iterations on the RT03 development set. The seed model (iteration 0) is initialized w/ or w/o online PL.} 
	\label{fig:two_step}
\end{figure}

Regarding offline PL, using LM to produce better pseudo-labels is helpful (shown in \autoref{tab:offline_decode}). Nonetheless, employing online PL to generate a better seed model for offline PL, i.e., two-step PL, leads to better performance. Moreover, as shown in \autoref{fig:two_step}, merely increasing the number of iterations for offline PL cannot continue to reduce the WER after a certain iteration, and it cannot match the performance of two-step PL.

Therefore, we can safely conclude that online PL and offline PL are complementary and could be combined to achieve better performance.

\section{Discussions}

In this section, we further discuss the proposed \method approach and the possible future extensions.

\begin{itemize}
    \item \emph{Extending to other UDA settings:} We assume access to the target style text corpus and development set. But they might not be satisfied in some scenarios, thus requiring the modifications of the \method approach. When the target style text corpus is unavailable, the source LM can be used instead in the offline PL. And when the development set is missing, UCF can degenerate to the confidence filtering approach, where $\gamma$ and $\eta$ are both $0$ as they can't be automatically estimated on the development set.
    \item \emph{Extending to other E2E-ASR models:} We demonstrated the effectiveness of the proposed approaches on CTC models in this work. But the proposed approaches are not model-specific and can be naturally extended to other E2E-ASR models. A future direction is to apply the idea of DPL, UCF and two-step PL on AED~\cite{chan2016listen} and RNN-T~\cite{graves2012sequence} models with some possible modifications.
    \item \emph{Extending to the model compression task:} We focused on improving the cross-domain performance and did not pay attention to model compression. Nonetheless, simple modifications of the \method can extend to the model compression task: replace the student model with a smaller model in the last iteration of offline PL or conduct model compression to the final model of \method.
    \item \emph{Improving the confidence filtering for online PL:} Since the auxiliary forward computation of UCF would significantly increase the training time of online PL, UCF is only applied in the offline PL and we did not modify UCF to adapt the online PL scenario. However, the success of UCF on offline PL proves that a better confidence estimation can improve pseudo-labeling performance. Therefore, a future direction is to improve the confidence estimation in online PL without dramatically increasing the training time.
    \item \emph{Integration with other complementary approaches:} \method concentrates on the self-supervision based approaches to improve UDA performance. An intuitive complementary approach is target domain data synthesis, as more in-domain data is likely beneficial. However, the target data synthesis approach would complicate the training recipe and is less correlated with the proposed approach in this work. Therefore, we leave it as a future direction to be explored.
\end{itemize}

\section{Conclusion}
\label{sec:conclusion}
In this work, we propose a systematic UDA approach \method to utilize self-supervision for cross-domain speech recognition of the unlabeled target data. \method is built based on SSL and PL in the pre-training and fine-tuning paradigm. It could effectively alleviate both pre-training and fine-tuning mismatch in the UDA scenario. 

On the one hand, to deal with the pre-training mismatch, we dive into the continued pre-training and data replay techniques for better pre-training adaptation without the bother of catastrophic forgetting. We show that simultaneous fine-tuning on both domains could alleviate the trouble of catastrophic forgetting. And data replay would be significantly helpful if we only fine-tune the pre-trained model on labeled source data. 

On the other hand, a domain adaptive fine-tuning approach is proposed to resolve a fine-tuning mismatch with PL, consisting of three unique techniques. Firstly, DPL alleviates the error accumulation of online PL; Secondly, UCF could select a better subset of pseudo-labels during offline PL; Lastly, the two-step PL improves the quality of pseudo-labels with LM.

Extensive ablation experiments are conducted to verify each component of the proposed approach. Experimental results demonstrate that \method significantly outperforms previous UDA approaches and their naive combinations. Furthermore, \method could achieve comparable performance with supervised training under minor domain mismatches.

\section*{Acknowledgment}

This work is partially supported by the National Key Research and Development Program of China (No. 2020AAA0108002), the Youth Innovation Promotion Association, Chinese Academy of Sciences, and the Frontier Exploration Project Independently Deployed by Institute of Acoustics, Chinese Academy of Sciences under Grant QYTS202011. 

\ifCLASSOPTIONcaptionsoff
  \newpage
\fi

\bibliographystyle{IEEEtran}
\bibliography{./main}

\end{document}